\documentclass[conference]{IEEEtran}
\IEEEoverridecommandlockouts
\usepackage{algorithm}                %
\usepackage{acronym}
\usepackage{algorithmic}              %
\usepackage{amsmath,amssymb,amsfonts} %
\usepackage{booktabs}                 %
\usepackage{cite}                     %
\usepackage{enumitem}%
\usepackage{graphicx}                 %
\usepackage{mathtools}                %
\usepackage{mathrsfs}                 %
\usepackage{pdfsync}  %
\usepackage{xcolor}                   %

\usepackage{tikz,pgfplots}
\pgfplotsset{compat=newest}
\usetikzlibrary{plotmarks}  %
\usetikzlibrary{matrix} %
\usepgfplotslibrary{external}
\newlength\figureheight
\newlength\figurewidth

\def\BibTeX{{\rm B\kern-.05em{\sc i\kern-.025em b}\kern-.08em
    T\kern-.1667em\lower.7ex\hbox{E}\kern-.125emX}}
\ifCLASSOPTIONcompsoc\usepackage[caption=false,font=normalsize,labelfont=sf,textfont=sf]{subfig}\else\usepackage[caption=false,font=footnotesize]{subfig}\fi

\usepackage[bookmarks=false]{hyperref} %

\newcommand{\gauss}[3]{\mathcal{N}(#1; \, #2, \, #3)}

\newcommand\varlist{,\!\makebox[1em][c]{.\hfil.\hfil.},}
\newcommand\prodvarlist{\!\makebox[1em][c]{$\cdot$\hfil$\cdot$\hfil$\cdot$}}

\DeclareMathOperator*{\argmax}{arg\,max}

\newtheorem{theorem}{Theorem}%
\newtheorem{proposition}[theorem]{Proposition}

\acrodef{cbmember}[CBMeMBer]{cardinality-balanced multi-object multi-Bernoulli}
\acrodef{cphd}[CPHD]{cardinalized probability hypothesis density}
\acrodef{cs}[CS]{Cauchy-Schwarz}
\acrodef{dddGLMB}[DD$\delta$-GLMB]{data-driven $\delta$-generalized labeled multi-Bernoulli}
\acrodef{dglmb}[$\delta$-GLMB]{$\delta$-generalized  labeled multi-Bernoulli}
\acrodef{ekf}[EKF]{extended Kalman filter}
\acrodef{em}[EM]{expectation maximization}
\acrodef{fisst}[FISST]{finite set statistics}
\acrodef{fov}[FoV]{field-of-view}
\acrodefplural{fov}[FoVs]{fields-of-view}
\acrodef{gm}[GM]{Gaussian mixture}
\acrodef{glmb}[GLMB]{generalized labeled multi-Bernoulli}
\acrodef{gospa}[GOSPA]{generalized optimal sub-pattern assignment}
\acrodef{iid}[i.i.d.]{independently and identically distributed}
\acrodef{iidc}[i.i.d.c.]{independently and identically distributed cluster}
\acrodef{kl}[KL]{Kullback-Leibler}
\acrodef{kld}[KLD]{Kullback-Leibler divergence}
\acrodef{lmb}[LMB]{labeled multi-Bernoulli}
\acrodef{lrfs}[LRFS]{labeled random finite set}
\acrodef{mb}[MB]{multi-Bernoulli}
\acrodef{mbm}[MBM]{multi-Bernoulli mixture}
\acrodef{mcmc}[MCMC]{Markov chain Monte Carlo}
\acrodef{mht}[MHT]{multiple hypothesis tracking}
\acrodef{mmse}[MMSE]{minimum mean square error}
\acrodef{mprint}[M-PrNTT]{multi-Bernoulli probabilistic new target tracker}
\acrodef{mtt}[MTT]{multitarget tracking}
\acrodef{pdf}[pdf]{probability density function}
\acrodef{pf}[PF]{particle filter}
\acrodef{pmf}[pmf]{probability mass function}
\acrodef{phd}[PHD]{probability hypothesis density}
\acrodefplural{phd}[PHDs]{probability hypothesis densities}
\acrodef{rfs}[RFS]{random finite set}
\acrodefplural{rfs}[RFSs]{random finite sets}
\acrodef{roi}[ROI]{region of interest}
\acrodefplural{roi}[ROIs]{regions of interest}
\acrodef{rss}[RSS]{root-sum-square}
\acrodef{rv}[RV]{random variable}

\colorlet{change1}{black}
\begin{document}

\title{%
Split Happens! Imprecise and Negative Information in Gaussian Mixture Random Finite Set Filtering\\
\thanks{Keith A. LeGrand is with the Sensing, Controls, and Probabilistic Estimation (SCOPE) Laboratory, School of Aeronautics and Astronautics, Purdue University, West Lafayette, Indiana, United States. Silvia Ferrari is with the Laboratory for Intelligent Systems and Controls (LISC), Sibley School of Mechanical and Aerospace Engineering, Cornell University, Ithaca, New York, United States. This work was supported in part by Office of Naval Research Grant N0014-19-1-2266 and by the National Defense Science and Engineering Graduate (NDSEG) Fellowship.}
}

\author{\IEEEauthorblockN{Keith A. LeGrand and Silvia Ferrari}
}

\maketitle

\begin{abstract}
In object tracking and state estimation problems, ambiguous evidence such as imprecise measurements and the absence of detections can contain valuable information and thus be leveraged to further refine the probabilistic belief state.
In particular, knowledge of a sensor's bounded field-of-view can be exploited to incorporate evidence of where an object was not observed.
This paper presents a systematic approach for incorporating knowledge of the field-of-view geometry and position and object inclusion/exclusion evidence into object state densities and random finite set multi-object cardinality distributions.
The resulting state estimation problem is nonlinear and solved using a new Gaussian mixture approximation based on recursive component splitting.
Based on this approximation, a novel Gaussian mixture Bernoulli filter for imprecise measurements is derived and demonstrated in a tracking problem using only natural language statements as inputs.
This paper also considers the relationship between bounded fields-of-view and cardinality distributions for a representative selection of multi-object distributions, which can be used for sensor planning, as is demonstrated through a problem involving a multi-Bernoulli process with up to one-hundred potential objects.
\end{abstract}


\section{Introduction}
\label{sec:Introductions}
Random finite set (RFS)\acused{rfs} theory has been proven a highly effective framework for developing and analyzing tracking and sensor planning algorithms in applications involving an unknown number of multiple targets (objects) \cite{MahlerStatisticalMultitargetFusion07, VoLabeledRfsGlmbFilter14, ReuterLmbFilter14, Garcia-fernandezGaussianImplementationPmbm19,HoangSensorManagementMultiBernoulliRenyiMaxCardinalityVariance14,BeardVoidProbabilitiesCauchySchwarzGlmb17,WangMultiSensorControlLmb18}.
Until recently, however, little attention has been devoted to the role that bounded sensor \acp{fov} play in assimilating measurements, or lack thereof, into multi-object probability distributions.
Existing tracking algorithms, for example, typically terminate object tracks when the object leaves the sensor \ac{fov}.
While this approach is suitable when the \ac{fov} doubles as the tracking \ac{roi}, it is inapplicable when the sensor \ac{fov} is much smaller than the \ac{roi} and, thus, must be moved or positioned so as to maximize information value \cite{FerrariGeometricDetectIntercept09, WeiGeometricTransversalsSensorPlanning15, GehlySearchDetectTrack18,BuonviriSurveyMultiSensorRfs19,legrand2022CellMultiBernoulliCellMBa,LegrandRandomFiniteSetSensorControl21}.
Other technical challenges arise in multi-sensor fusion problems involving bounded overlapping \acp{fov} and have been the focus of recent work\cite{li2019LocalDiffusionBasedDistributedSMCPHD,da2020GaussianMixtureParticle,gao2022FusionLmbMdglmbDifferentFov,wang2022MultiAgentFusionBird}.

The simple indication of an object's presence or absence within a known region, such as an \ac{fov},  is powerful evidence that can be incorporated to update the object \ac{pdf} in a Bayesian framework.
For example, the absence of detections is a type of \textit{negative information} indicating that the object state may reside outside the \ac{fov} \cite{KochExploitingNegativeEvidence07}.
In contrast, binary-type sensors may produce \emph{imprecise measurements} \cite{GoodmanMathematicsOfDataFusion97,GningBernoulliParticleFilterTripleUncertainty12,RisticParticleFiltersBook13} that indicate the object is inside the sensor \ac{fov} but provide no further localization information.
Similarly, ``soft'' data from human sources, such as natural language statements, can be considered as imprecise measurements due to their inherent ambiguity \cite{BishopFusionNaturalLanguageRandomSet13,RisticBernoulliFilterTutorial13}.
Particle-based filtering algorithms \cite{ArulampalamParticleFilterTutorial02,VoSequentialMonteCarlo05,RisticParticleFiltersBook13} can accommodate such measurements but require a large number of particles and are computationally expensive.
The integrated track splitting filter for state-dependent probability of detection (ITSpd) \cite{SongTargetTrackingStateDependentPd11} uses \acp{gm} to model both the object \ac{pdf} and the state-dependent probability of detection function.
Though \acp{gm} efficiently model some detection probability functions, other simple functions, such as uniform probability densities over a 3D \ac{fov}, require problematically large numbers of components.
Other approaches \cite{AhmedSoftDataFusionVariationalBayesGm13,WeiDistributedSpaceTrackingFovEm18} employ stochastic sampling and the \ac{em} algorithm to compute \ac{gm} approximations to the posterior \ac{pdf}.
However, the use of intermediate particle representations and \ac{em} reconstruction can lead to information loss, and  convergence is sensitive to \ac{em} initial condition specification.

\begin{figure*}[htpb]
  \centering
  \includegraphics[height=0.3\linewidth]{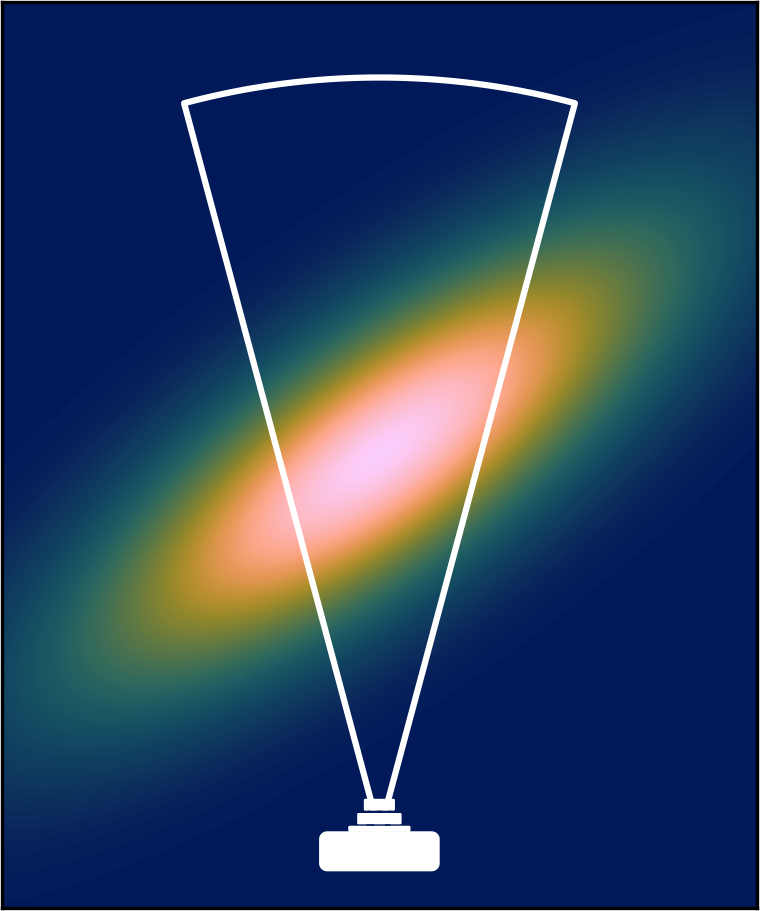}
  \includegraphics[height=0.3\linewidth]{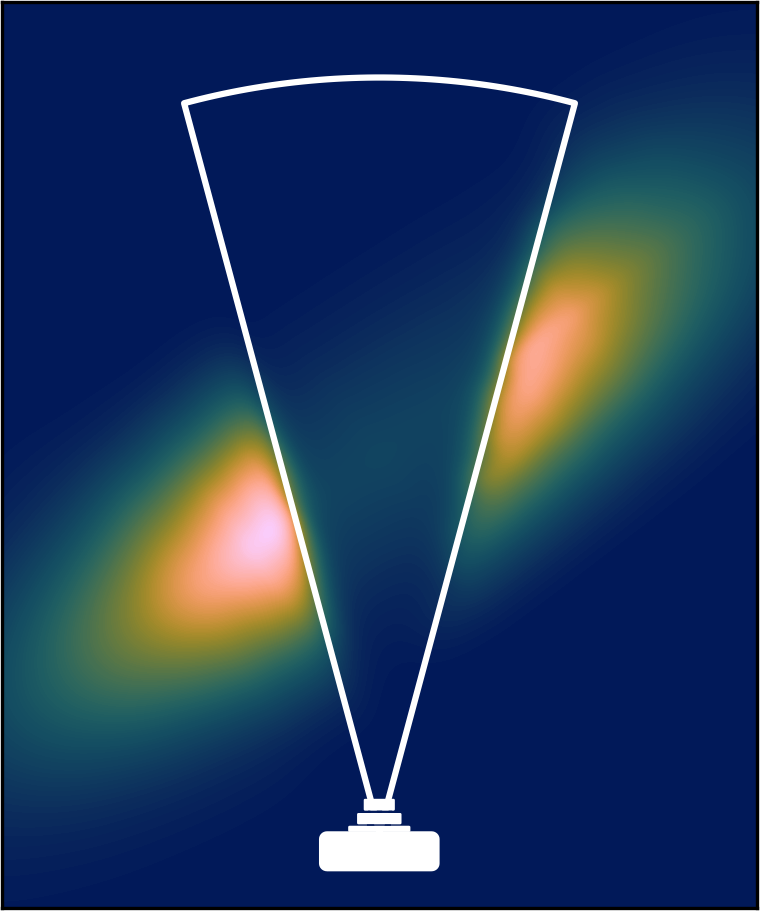}\hfill
  \includegraphics[height=0.3\linewidth]{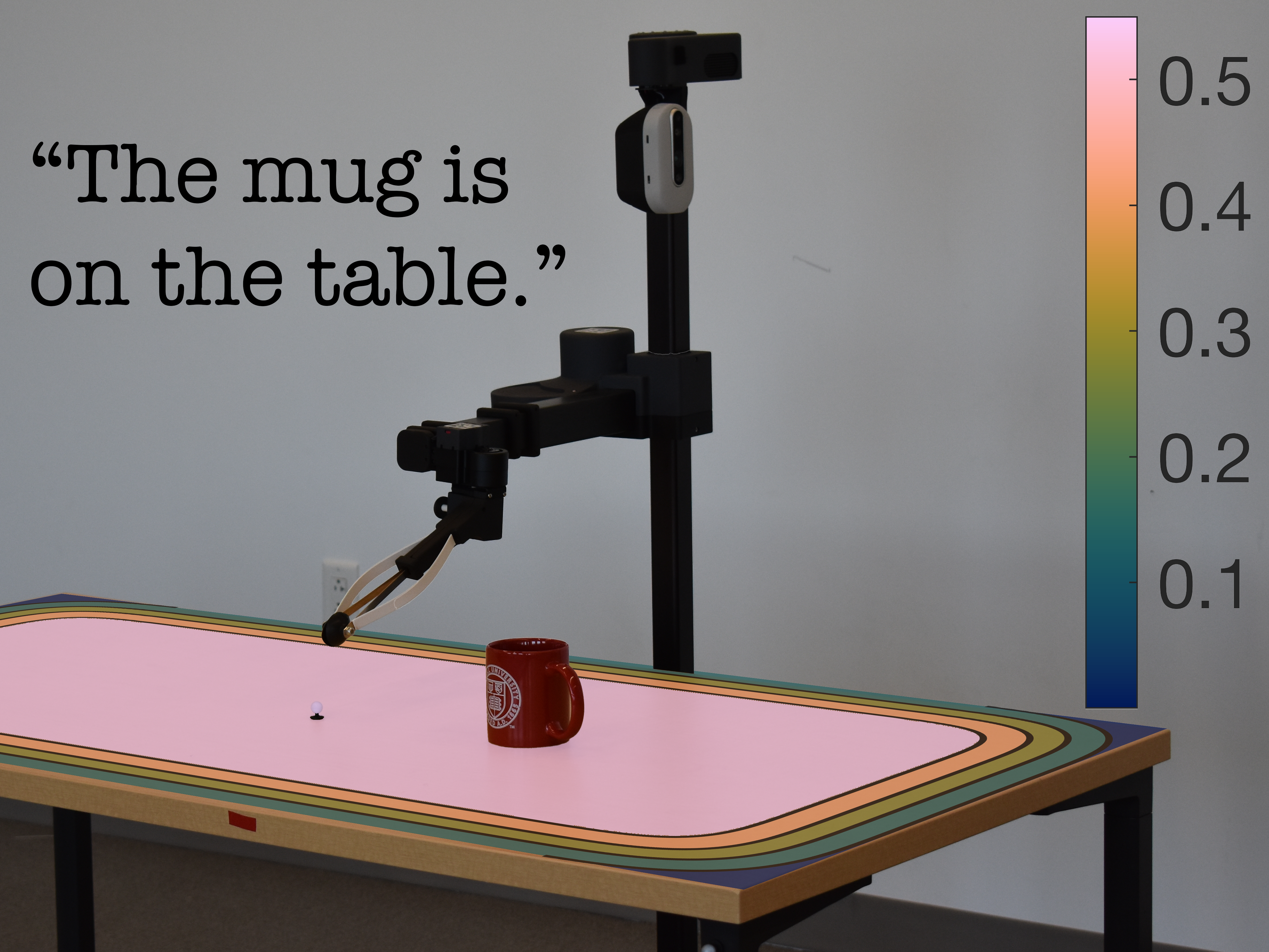}\,
  \caption{Gaussian mixture probability distribution before (left) and after (center) incorporating negative information (that is, the absence of detections) and the known bounded sensor field-of-view, and a Gaussian mixture distribution after incorporating an imprecise measurement corresponding to a set of possible mug locations (right) in a robot perception application.}%
  \label{fig:MugOnTable}
\end{figure*}

This paper presents new methods for incorporating inclusion/exclusion evidence in Bayesian single- and multi-object estimation and sensor planning algorithms, as illustrated in Figure~\ref{fig:MugOnTable}.

Section~\ref{sec:background_and_notation} defines the notation used in this paper, and Section~\ref{sec:ProblemFormutation} details the problem formulation and related assumptions.
Section~\ref{sec:GmPartitioning} presents a deterministic method that partitions a \ac{gm} state density along the boundaries of a known region, such as an \ac{fov}, through recursive Gaussian splitting.
By this approach, negative information is leveraged in \ac{gm} filters to further refine the posterior object state \ac{pdf}.
Similarly, imprecise measurements, such as natural language statements, can be incorporated to obtain \ac{gm} posterior distributions using a new multi-\ac{fov}-generalized splitting algorithm.
Section~\ref{sec:application_to_imprecise_measurements} presents an application of the splitting method to the tracking of a person in a crowded space using natural language statements and a new \ac{gm} Bernoulli filter algorithm.
In Section~\ref{sec:FovCardinality}, \ac{fov} object cardinality \acp{pmf} are derived for some of the most commonly encountered \ac{rfs} distributions.
Section~\ref{sec:SensorPlacementExample} presents an application of bounded \ac{fov} statistics to a sensor placement problem, and conclusions are made in Section~\ref{sec:Conclusions}.
This paper builds on previous work, \cite{LeGrandRoleOfBoundedFovs20}, by presenting a generalized partitioning algorithm for use with multiple \acp{fov}, a derivation of a new \ac{gm} Bernoulli filter algorithm applicable to imprecise measurements, and a simulation of a tracking problem using natural language statements.

\section{Notation}%
\label{sec:background_and_notation}

Throughout this paper, single-object states are represented by lowercase letters~(e.g.~$\mathbf{x}$, $\mathring{x}$), while multi-object states are represented by italic uppercase letters~(e.g.~$X$,~$\mathring{X}$).  Bold lowercase letters are used to denote vectors and bold uppercase letters are used denote matrices.
The accent ``$\mathring{\,\,\,}$'' is used to distinguish labeled states and functions (e.g.~$\mathring{f}$,~$\mathring{x}$,~$\mathring{X}$) from their unlabeled equivalents (e.g.~$f$,~$x$,~$X$).
Spaces are represented by blackboard bold symbols~(e.g.~$\mathbb{X}$,~$\mathbb{L}$).

The multi-object exponential notation,
\begin{align}
  h^{A} \triangleq \prod_{a \in A} h(a)
\end{align}
where $h^{\emptyset}\triangleq 1$, is adopted throughout.
For multivariate functions, the dot $(\cdot)$ denotes the argument of the multi-object exponential, e.g.:
\begin{align}
  [g(a, \cdot, c)]^{B}
  \triangleq
  \prod_{b \in B}g(a, b, c)
\end{align}
The exponential notation is used to denote the product space, $\mathbb{X}^{n} = \prod^{n} (\mathbb{X} \times)$, whereas exponents of \acp{rfs} are used to denote \acp{rfs} of a given cardinality, e.g.\ $|X^{n}| = n$, where $n$ is the cardinality.
The set of natural numbers less than or equal to $n$ is denoted by
\begin{align}
  \mathbb{N}_{n} \triangleq \{1, \hdots, n\}
\end{align}
The operator $\textrm{diag}(\cdot)$ places its input on the diagonal of the zero matrix.  The Kronecker delta function is defined as
\begin{align}
  \delta_{\boldsymbol{a}}(\boldsymbol{b}) \triangleq
  \left\{\begin{array}{ll}
  1, & \text{if } \boldsymbol{b} = \boldsymbol{a} \\
  0, & \text{otherwise}
  \end{array} \right.
\end{align}
for any two arbitrary vectors $\boldsymbol{a}, \,\boldsymbol{b} \in \mathbb{R}^{n}$.  The inner product  of two integrable functions $f(\cdot)$ and $g(\cdot)$ is denoted by
\begin{align}
  \left<f,g\right> = \int f(\mathbf{x})g(\mathbf{x})\mathrm{d}\mathbf{x}
\end{align}
\section{Problem Formulation and Assumptions}
\label{sec:ProblemFormutation}
This paper considers the incorporation of inclusion/exclusion evidence into algorithms for \mbox{(multi-)object} tracking and sensor planning when the number of objects is unknown and time-varying.
Often in tracking, object detection may depend only on a partial state $\mathbf{s}\in\mathbb{X}_{s} \subseteq \mathbb{R}^{n_{s}}$, where $\mathbb{X}_{s} \times \mathbb{X}_{v} = \mathbb{X}\subseteq \mathbb{R}^{n_{x}}$ forms the full object state space.
For example, the instantaneous ability of a sensor to detect an object may depend only on the object's relative position.
In that case, $\mathbb{X}_{s}$ is the position space, and $\mathbb{X}_{v}$ is comprised of non-position states, such as object velocity.
This nomenclature is adopted throughout the paper, while noting that the approach is applicable to other state definitions.
Following \cite{BaumgartnerOptimalControlUnderwaterSensor09}, the sensor \ac{fov} can be defined as the compact subset $\mathcal{S}(\boldsymbol{q})\subset\mathbb{X}_{s}$.
In general, the \ac{fov} is a function of the sensor state $\boldsymbol{q}$, which, for example, may consist of the sensor position, orientation, and zoom level.
However, for notational simplicity this dependence is omitted in the remainder of this paper.

Now, let the object state $\mathbf{x}$ consist of the kinematic variables that are to be estimated from data via filtering, such as the object position, velocity, and turn rate.  Then, the single-object \ac{pdf} is denoted by $p(\mathbf{x})$.  Letting $\mathbf{s}=\textrm{proj}_{\mathbb{X}_{s}} \mathbf{x}$ denote the state elements that correspond to $\mathbb{X}_{s}$, an object's presence inside the \ac{fov} can be expressed by the generalized indicator function
\begin{align}
  1_{\mathcal{S}}(\mathbf{x}) = \left\{\begin{array}{ll}
      1, & \text{if } \mathbf{s} \in \mathcal{S} \\
  0, & \text{otherwise}
  \end{array} \right.
\end{align}
The number of objects and their kinematic states are unknown \emph{a priori}, but can be assumed to consist of discrete and continuous variables, respectively.  The collection of object states is modeled as an \ac{rfs} $X$ or \ac{lrfs} $\mathring{X}$, where the single-object labeled state $\mathring{x} = (\mathbf{x}, \ell)\in \mathbb{X} \times \mathbb{L}$ consists of a kinematic state vector $\mathbf{x}$ and unique discrete label $\ell$.  It is assumed that the prior multi-object distribution is known, e.g., from the output of a multi-object filter, and modeled using either the \ac{rfs} density $f(X)$ or \ac{lrfs} density~$\mathring{f}(\mathring{X})$.

In \ac{rfs}-based tracking, single-object densities are, in fact, parameters of the higher-dimensional multi-object density.  Non-Gaussian single-object state densities are often modeled using \acp{gm} because they admit closed-form approximations to the multi-object Bayes recursion under certain conditions \cite{VoLabeledRfsGlmbFilter14, VoGaussianMixturePhd06}.  Therefore, in this paper, it is assumed that single-object densities (which are parameters of the higher dimensional multi-object density) are parameterized as
\begin{align}
  p(\mathbf{x}) = \sum_{\ell=1}^{L} w^{(\ell)} \gauss{\mathbf{x}}{\mathbf{m}^{(\ell)}}{\mathbf{P}^{(\ell)}}
\end{align}
where $L$ is the number of \ac{gm} components and $w^{(\ell)}$, $\mathbf{m}^{(\ell)}$, and $\mathbf{P}^{(\ell)}$ are the weight, mean, and covariance matrix of the $\ell$\textsuperscript{th} component, respectively.

In this paper, the problem considered is forming \ac{gm} Bayesian posteriors given evidence of the forms:
\begin{enumerate}[label=T\arabic*.]
  \item \textit{The existence or non-existence of a measurement is evidence of the inclusion or exclusion of the object state within a known set.} For example, the non-existence of a detection (measurement) is evidence of an object's position exclusion from the sensor \ac{fov}.
  \item \textit{The value of the measurement is evidence of the inclusion or exclusion of the object state within a known set}. For example, the observation that a sea-level freshwater lake is frozen is evidence that the water temperature belongs to the set of temperatures below $0\,^{\circ} \textrm{C}$.
\end{enumerate}

Mahler's \ac{fisst} provides the mathematical machinery for modeling types T1 and T2 using state-dependent probability of detection functions and generalized likelihood functions, respectively.
However, in both cases, the Bayes posterior involves products of the prior \ac{gm} with indicator functions such as
\begin{align}
  \label{eq:1sp}
  p(\mathbf{x})1_{\mathcal{S}}(\mathbf{x}) &\triangleq p_{\mathcal{S}}(\mathbf{x})
  \qquad \textrm{and}\\
  \label{eq:1m1sp}
  (1 - 1_{\mathcal{S}}(\mathbf{x})) p(\mathbf{x})  &\triangleq p_{\mathcal{C}(\mathcal{S})}(\mathbf{x})
\end{align}
where $\mathcal{C}(\mathcal{S})$ denotes the complement space $\mathbb{X}_{s} \setminus \mathcal{S}$.
Thus, the resulting posterior is no longer a \ac{gm}.

This paper presents a fast \ac{gm} approximation of (\ref{eq:1sp}) and (\ref{eq:1m1sp}), thereby enabling the assimilation of inclusion/exclusion evidence in any \ac{gm}-based \ac{rfs} single-object or multi-object filter.
Building on these concepts, this paper also considers the role of inclusion/exclusion evidence in object cardinality distributions and derives \ac{pmf} expressions that describe the probabilities associated with different numbers of objects existing within a given set $\mathcal{S}$ (such as an \ac{fov}).
\section{GM Approximation of \ac{fov}-partitioned Densities}
\label{sec:GmPartitioning}
This section presents a method for partitioning the object \ac{pdf} into truncated densities $p_{\mathcal{S}}(\mathbf{x})$ and $p_{\mathcal{C}(\mathcal{S})}(\mathbf{x})$, with supports $\mathcal{S} \times \mathbb{X}_{v}$ and $\mathcal{C}(\mathcal{S}) \times \mathbb{X}_{v}$, respectively.

Focus is given to the single-object state density with the awareness that the method is naturally extended to \ac{rfs} multi-object densities and algorithms that use \ac{gm} parameterization. Consider the single-object density $p(\mathbf{x})$ parameterized by an $L$-component \ac{gm}, as follows:
\begin{align}
  p(\mathbf{x})
  =
  p_{\mathcal{S}}(\mathbf{x}) + p_{\mathcal{C}(\mathcal{S})}(\mathbf{x})
  =
  \sum_{\ell=1}^{L} w^{(\ell)} \mathcal{N}(\mathbf{x}; \, \mathbf{m}^{(\ell)}, \mathbf{P}^{(\ell)})
\end{align}
One simple approximation of densities partitioned according to the discrete \ac{fov} geometry, referred to as \ac{fov}-partitioned densities hereon, is found by evaluating the indicator function at the component means \cite{LegrandRelativeSpaceObjectTrackingFusion15}, i.e.:
\begin{align}
  \label{eq:InSTimesPApproxGm}
  p_{\mathcal{S}}(\mathbf{x})
  &\approx
  \sum_{\ell=1}^{L} w^{(\ell)}
  1_{\mathcal{S}}(\mathbf{m}^{(\ell)})
  \gauss{\mathbf{x}}{\mathbf{m}^{(\ell)}}{\mathbf{P}^{(\ell)}}\\
  \label{eq:NotInSTimesPApproxGm}
  p_{\mathcal{C}(\mathcal{S})}(\mathbf{x})
  &\approx
  \sum_{\ell=1}^{L} w^{(\ell)}
  (1 - 1_{\mathcal{S}}(\mathbf{m}^{(\ell)}))
  \gauss{\mathbf{x}}{\mathbf{m}^{(\ell)}}{\mathbf{P}^{(\ell)}}
\end{align}
By this approach, components whose means lie inside (outside) the \ac{fov} are preserved (pruned), or vice versa.

The accuracy of this mean-based partition approximation depends strongly on the resolution of the \ac{gm} near the geometric boundaries of the \ac{fov}.  Even though the mean of a given component lies inside (outside) the \ac{fov}, a considerable proportion of the probability mass may lie outside (inside) the \ac{fov}, as is illustrated in Figure~\ref{fig:Example2d}a.  Therefore, the amount of \ac{fov} overlap, along with the weight of the component, determines the accuracy of the approximations~(\ref{eq:InSTimesPApproxGm})-(\ref{eq:NotInSTimesPApproxGm}).  To that end, the algorithm presented in the following subsection iteratively resolves the \ac{gm} near \ac{fov} bounds by recursively splitting Gaussian components that overlap the \ac{fov} bounds.
\newlength\gaussquadwidth
\setlength\gaussquadwidth{4.2cm}
\begin{figure}[h!]
  \centering
  \subfloat[]{\includegraphics[height=\gaussquadwidth]{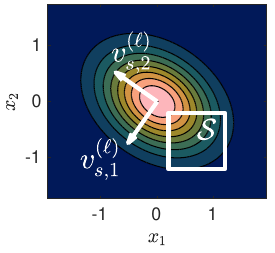}}\hfil
  \subfloat[]{\includegraphics[height=\gaussquadwidth]{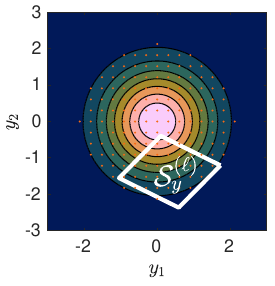}}%
  \caption{Original component density and FoV with covariance eigenvectors overlaid (a), and same component density and FoV after change of variables (b).}
  \label{fig:Example2d}
\end{figure}

\subsection{Gaussian Splitting Algorithm}
The Gaussian splitting algorithm presented in this subsection forms an \ac{fov}-partitioned \ac{gm} approximation of the original \ac{gm} by using a higher number of components near the \ac{fov} boundaries, $\partial \mathcal{S}$, so as to improve the accuracy of the mean-based partition.

Consider for simplicity a two-dimensional example in which the original \ac{gm}, $p(\mathbf{x})$, has a single component whose mean lies outside the \ac{fov}, as shown in Figure~\ref{fig:Example2d}a.
The algorithm first applies a change of variables $\mathbf{x} \mapsto \mathbf{y}\in \mathbb{Y}\subseteq \mathbb{R}^{n_{s}}$ such that $p(\mathbf{y})$ is symmetric and has zero mean and unit variance.
The basis vectors of the space $\mathbb{Y}$ correspond to the principal directions of the component's position covariance.
The same change of variables is applied to the \ac{fov} bounds (Figure~\ref{fig:Example2d}b).

A pre-computed point grid is then tested for inclusion in the transformed \ac{fov} in order to decide whether to split the component, and if so, along which principal direction.
For each new split component, the process is repeated--if a new component significantly overlaps the \ac{fov} boundaries, it may be further split into several smaller components, as illustrated in Figure~\ref{fig:Example2dSplitContours}.
This process is repeated until stopping criteria are satisfied.
After the \ac{gm} splitting terminates, $p_{\mathcal{S}}(\mathbf{x})$ and $p_{\mathcal{C}(\mathcal{S})}(\mathbf{x})$ are approximated by the mean-based partition (Eqs.~\ref{eq:1sp}-\ref{eq:1m1sp}), as illustrated in Figure~\ref{fig:Example2dSplitInOutFov}.
\setlength\gaussquadwidth{3.7cm}
\begin{figure}[htbp]
  \centering
  \subfloat[]{\includegraphics[height=\gaussquadwidth]{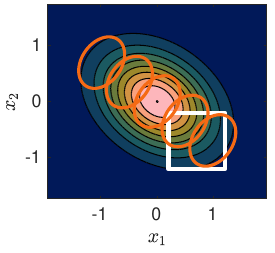}}\hfil
  \subfloat[]{\includegraphics[height=\gaussquadwidth]{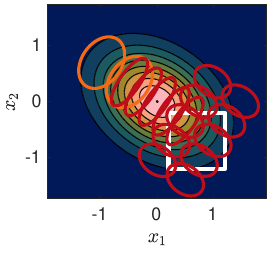}}
  \caption{$1\sigma$ contours of components after first split operation (a), and second split operation (b), where components formed in the second operation are shown in red.}
  \label{fig:Example2dSplitContours}
\end{figure}
\begin{figure}[bhtp]
  \centering
  \subfloat[]{\includegraphics[height=\gaussquadwidth]{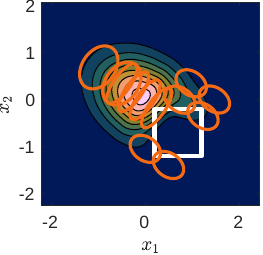}}\hfil
  \subfloat[]{\includegraphics[height=\gaussquadwidth]{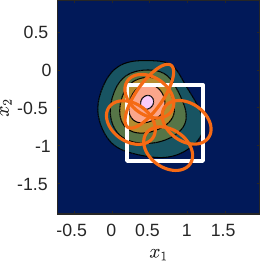}}%
  \caption{The densities $p_{\mathcal{C}(\mathcal{S})}(\mathbf{x})$ (a), and $p_{\mathcal{S}}(\mathbf{x})$ (b), which have been approximated using two iterations of component splitting and the subsequent mean-based partition.}
  \label{fig:Example2dSplitInOutFov}
\end{figure}

\subsection{Univariate Splitting Library}
Splitting is performed efficiently by utilizing a pre-generated library of optimal split parameters for the univariate standard Gaussian $q(x)$, as first proposed in \cite{HuberEntropyApproximationSplitLibrary08} and later generalized in \cite{DemarsEntropyPropagationGaussSplit13}.  The univariate split parameters are retrieved at run-time and applied to arbitrary multivariate Gaussian densities via scaling, shifting, and covariance diagonalization.

Generation of the univariate split library is performed by minimizing the cost function
\begin{align}
  J = L_{2}(q||\tilde{q}) + \lambda \tilde{\sigma}^{2} \qquad \textrm{s.t.} \sum_{j=1}^{R} \tilde{w}^{(j)}=1
\end{align}
where
\begin{align}
  \tilde{q}(x) = \sum \limits_{j=1}^{R} \tilde{w}^{(j)} \gauss{x}{\tilde{m}^{(j)}}{\tilde{\sigma}^{2}}
\end{align}
for different parameter values $R$, $\lambda$.  The regularization term $\lambda$ balances the importance of using smaller standard deviations $\tilde{\sigma}$ with the minimization of the $L_{2}$ distance. While other distance measures may be used, the $L_{2}$ distance is attractive because it can be computed in closed form for \acp{gm} \cite{DemarsEntropyPropagationGaussSplit13}.
As an example, the optimal split parameters for $R=4$, $\lambda=0.001$ are provided in Table~\ref{t:UnivariateSplitParameters}.
\begin{table}[htbp]
  \centering
  \caption{Univariate split parameters for $R=4$, $\lambda=0.001$.}
  \label{t:UnivariateSplitParameters}
  \begin{tabular}{rrrr}
    \toprule
    $j$ & $\tilde{w}^{(j)}$ & $\tilde{m}^{(j)}$ & $\tilde{\sigma}$ \\
    \hline
    $1$& $0.10766586425362$ & $-1.42237156603631$ & $0.58160633157686$ \\
    $2$& $0.39233413574638$ & $-0.47412385534547$ & $0.58160633157686$ \\
    $3$& $0.39233413574638$ & $ 0.47412385534547$ & $0.58160633157686$ \\
    $4$& $0.10766586425362$ & $ 1.42237156603631$ & $0.58160633157686$
   \end{tabular}
\end{table}

\subsection{Change of Variables}
The determination of which components should be split, and if so, along which direction, is simplified by first establishing a change of variables.  For each component with index $\ell$, the change of variables $\boldsymbol{h}^{(\ell)}: \mathbb{X}_{s} \mapsto \mathbb{Y}$ is applied as follows:
\begin{align}
  \label{eq:TransformationToZ}
  \mathbf{y}
  =
  \boldsymbol{h}^{(\ell)}(\mathbf{s}; \mathbf{m}_{s}^{(\ell)}, \mathbf{P}_{s}^{(\ell)})
  \triangleq
  (\boldsymbol{\Lambda}_{s}^{(\ell)})^{-\frac{1}{2}}\boldsymbol{V}_{s}^{(\ell)T} (\mathbf{s} - \mathbf{m}_{s}^{(\ell)})
\end{align}
where
\begin{align}
  \boldsymbol{V}_{s}^{(\ell)}
  &=
  [\boldsymbol{v}_{s,1}^{(\ell)} \quad \cdots \quad \boldsymbol{v}_{s,n_{s}}^{(\ell)}]\\
  (\boldsymbol{\Lambda}_{s}^{(\ell)})^{-1/2}
  &=
  \textrm{diag}
  \left(
    \left[
      \tfrac{1}{\sqrt{\lambda_{s, 1}^{(\ell)}}}\quad  \cdots \quad \tfrac{1}{\sqrt{\lambda_{s,n_{s}}^{(\ell)}}}
    \right]
  \right)
\end{align}
and $\mathbf{m}_{s}^{(\ell)}$ is the $n_{s}$-element position portion of the full-state mean, and the columns of $\boldsymbol{V}_{s}^{(\ell)}$ are the normalized eigenvectors of the position-marginal covariance $\mathbf{P}_{s}^{(\ell)}$, with $\boldsymbol{v}_{s, i}^{(\ell)}$ corresponding to the $i$\textsuperscript{th} eigenvalue $\lambda_{s, i}^{(\ell)}$.  In the transformed space,
\begin{align}
  p(\mathbf{y}) = \gauss{\mathbf{y}}{\boldsymbol{0}}{\boldsymbol{I}}
\end{align}
Note that, in defining the transformation over $\mathbb{X}_{s}$, the same transformation can be applied to the \ac{fov}, such that
\begin{align}
  \label{eq:Sz}
  \mathcal{S}_{y}^{(\ell)} = \{\boldsymbol{h}^{(\ell)}(\mathbf{s}; \mathbf{m}_{s}^{(\ell)}, \mathbf{P}_{s}^{(\ell)}) : \mathbf{s} \in \mathcal{S}\}
\end{align}

In $\mathbb{Y}$, the Euclidean distances to boundary points of $\mathcal{S}_{y}^{(\ell)}$ can be interpreted as probabilistically normalized distances.  In fact, the Euclidean distance of a point $\mathbf{y}$ from the origin in $\mathbb{Y}$ corresponds exactly to the Mahalanobis distance between the corresponding point $\mathbf{s}$ and the original position-marginal component.

\subsection{Component Selection and Collocation Points}
Components are selected for splitting if they have sufficient weight and significant statistical overlap of the \ac{fov} boundaries ($\partial \mathcal{S}$).  For components of sufficient weight, the change of variables is applied to the \ac{fov} to obtain $\mathcal{S}^{(\ell)}_{y}$ per~(\ref{eq:Sz}).  The overlap of the original component on $\mathcal{S}$ is then equivalent to the overlap of the standard Gaussian distribution on $\mathcal{S}^{(\ell)}_{y}$, which is quantified using a grid of collocation points on $\mathbb{Y}$, as shown in Fig.~\ref{fig:Example2d}b.

Define the collocation point $\bar{\mathbf{y}}_{i_{1}, \dots, i_{n_{s}}} \in \mathbb{Y}$ such that
\begin{align}
  \bar{\mathbf{y}}_{i_{1}, \cdots, i_{n_{s}}}
  &\triangleq
  [\bar{y}_{1}(i_{1}) \, \dots \, \bar{y}_{n_{s}}(i_{n_{s}})]^{T},
  \quad
  (i_{1}, \dots, i_{n_{s}}) \in G\\
  \bar{y}_{j}(l)
  &=
  -\zeta + 2\zeta \left(\frac{l - 1}{N_{g}-1}\right),
  \qquad
  j\in \mathbb{N}_{n_{s}}\\
  G
  &=
  \{(i_{1} \varlist i_{n_{s}}) : i_{(\cdot)} \in \mathbb{N}_{N_{g}}, \, \|\mathbf{y}_{i_{1}, \hdots, i_{n_{s}}}\| \leq \zeta\}
\end{align}
where $\zeta$ is a user-specified bound for the grid, $G$ is the set of indices of points that are within $\zeta$ of the origin, and $N_{g}$ is the upper bound of the number of points per dimension.  An inclusion variable is defined as
\begin{align}
  \label{eq:ComponentInclusionVariable}
  d_{i_{1},\dots, i_{n_{s}}}^{(\ell)}
  \triangleq
  1_{\mathcal{S}_{y}^{(\ell)}}(\bar{\mathbf{y}}_{i_{1}, \dots, i_{n_{s}}})
\end{align}
Inclusion and exclusion patterns across the grid can be examined by first establishing an arbitrary reference index $(i_{1}', \dots, i_{n_{s}}')\in G$.
With this, $\varrho_{\mathcal{S}_{y}^{(\ell)}} \in \{0,1\}$ is established to mark total inclusion or total exclusion as
\begin{align}
  \label{eq:TotalInclusionExclusionSingleSensor}
  \varrho_{\mathcal{S}_{y}^{(\ell)}}
  =
  \prod_{G}{\delta_{d_{i_{1}', \dots, i_{n_{s}}'}^{(\ell)}}}
  (d_{i_{1}, \dots, i_{n_{s}}}^{(\ell)})
\end{align}
which is equal to unity if all grid points lie inside of $\mathcal{S}_{y}^{(\ell)}$ or all grid points lie outside of $\mathcal{S}_{y}^{(\ell)}$, and is zero otherwise.
If either all or no points are included, no splitting is required.  Otherwise, the component is marked for splitting.

\subsection{Position Split Direction}
\label{sec:position_split_direction}
Rather than split a component along each of its principal directions, a more judicious selection can be made by limiting split operations to a single direction (per component) per recursion.  Thus, by performing one split per component per recursion, the component selection criteria are re-evaluated, reducing the overall number of components generated.  In the aforementioned two-dimensional example, only a subset of new components generated from the first split are selected for further splitting as shown in Figure~\ref{fig:Example2dSplitContours}b.

The split direction is chosen based on the relative geometry of the \ac{fov}, and thus position vectors are of interest.
Choosing the best position split direction is a challenging problem.
A common approach is to split along the component's covariance eigenvector with the largest eigenvalue \cite{HuberEntropyApproximationSplitLibrary08}.
This strategy, however, does not consider the \ac{fov} geometry and thus may increase the mixture size without improvement to the \ac{fov}-partitioned densities (\ref{eq:InSTimesPApproxGm})-(\ref{eq:NotInSTimesPApproxGm}).
Ref.~\cite{HuberAdaptiveGaussianMixtureSplittingLinearization11} provides a more sophisticated split direction criterion based on the integral linearization errors along the covariance eigenvectors.
However, in the case that the \ac{fov} does not intersect the eigenvectors, this criterion cannot distinguish the best split direction.
Another approach \cite{TuggleAutomatedSplitting18} determines the split direction based on the Hessian of the underlying nonlinear transformation, evaluated at the component mean.
However, for the transformations considered in this paper of the form $g(\mathbf{s})=c\cdot1_{\mathcal{S}}(\mathbf{s})$, where $c$ is some arbitrary constant, the Hessian either vanishes (for $\mathbf{s}\notin \partial \mathcal{S}$) or is undefined (for $\mathbf{s}\in \partial \mathcal{S}$).

Ideally, splitting along the chosen direction should minimize the number of splits required in the next iteration as well as improve the accuracy of the partition approximation applied after the final iteration.
The computational complexity of exhaustive optimization of the split direction would likely negate the computational efficiency of the overall algorithm.
Instead, to minimize the number of splits required in the next iteration, the position split direction is chosen as the direction that is orthogonal to the most grid planes of consistent inclusion/exclusion.
Introducing a convenience function $s_{j}^{(\ell)}: \mathbb{N}_{N_{g}} \mapsto \{0,1\}$, the plane of constant $y_{j}=\bar{y}_{j}(l)$ is consistently inside or consistently outside if
\begin{align}
  \label{eq:ConsistentPlaneIndicator}
  s_{j}^{(\ell)}(l) = \prod_{G,i_{j}=l} {\delta_{d_{i_{1}',\dots, i_{j}, \dots, i_{n_{s}}'}^{(\ell)}}
  (d_{i_{1},\dots, i_{j}, \dots, i_{n_{s}}}^{(\ell)})}
\end{align}
is equal to unity, where $i_{1}',\dots, i_{j}, \dots, i_{n_{s}}'$ is an arbitrary index tuple in $G$ satisfying $i_{j}=l$, to which inclusion consistency is compared (see Appendix~\ref{sec:inclusion_consistency_example} for a numerical example). The optimal position split direction is then given by the eigenvector $\boldsymbol{v}_{s, j^{*}}$, where the optimal eigenvector index is found as
\begin{align}
  \label{eq:BestPositionSplitDirection}
  j^{*} = \argmax_{j} \left(\sum_{l=1}^{N_{g}} s_{j}^{(\ell)}(l)\right)
\end{align}
For notational simplicity, the implicit dependence of $j^{*}$ on the component index $\ell$ is omitted.
For example, referring back to the two-dimensional example and Figure~\ref{fig:Example2d}b, there are more rows than columns that are consistently inside or outside the transformed \ac{fov}, and thus $j^{*}=2$ is chosen as the desired position split direction index.  In the case where multiple maxima exist, the eigenvector with largest eigenvalue is selected, which corresponds to the direction of largest variance among the maximizing eigenvectors.

\subsection{Multivariate Split of Full-state Component}
Gaussian splitting must be performed along the principal directions of the full-state covariance.  The general multivariate split approximation, splitting along the $k$\textsuperscript{th} eigenvector $\boldsymbol{v}_{k}^{(\ell)}$ is given by \cite{DemarsEntropyPropagationGaussSplit13}
\begin{align}
  \label{eq:SplitMultivariate}
  w^{(\ell)} \gauss{\mathbf{x}}{\mathbf{m}^{(\ell)}}{\mathbf{P}^{(\ell)}} \approx \sum_{j=1}^{R} w^{(\ell,j)} \gauss{\mathbf{x}}{\mathbf{m}^{(\ell,j)}}{\mathbf{P}^{(\ell,j)}}
\end{align}
where
\begin{align}
  w^{(\ell,j)}
  &=
  \tilde{w}^{(j)} w^{(\ell)}\\
  \mathbf{m}^{(\ell, j)}
  &=
  \mathbf{m}^{(\ell)} + \sqrt{\lambda_{k}^{(\ell)}}
  \tilde{m}^{(j)} \boldsymbol{v}_{k}^{(\ell)}\\
  \mathbf{P}^{(\ell,j)}
  &=
  \boldsymbol{V}^{(\ell)} \boldsymbol{\Lambda}^{(\ell)} \boldsymbol{V}^{(\ell)T}\\
  \boldsymbol{\Lambda}^{(\ell)}
  &=
  \textrm{diag}
  \left([
    \lambda_{1} \, \cdots \, \tilde{\sigma}^{2} \lambda_{k} \, \cdots \, \lambda_{n_{x}}
  ]\right)
\end{align}
and the optimal univariate split parameters $\tilde{w}^{(j)}$, $\tilde{m}^{(j)}$, and $\tilde{\sigma}$ are found from the pre-computed split library given the number of split components $R$ and regularization parameter $\lambda$.  In general, the position components of the full-state eigenvectors will not perfectly match the desired position split vector due to correlations between the states.  Rather, the actual full-state split is performed along $\boldsymbol{v}_{k*}^{(\ell)}$, where the optimal eigenvector index is found according to
\begin{align}
  \label{eq:BestFullStateSplitDirection}
  k^{*} = \argmax_{k} \big|\big[\boldsymbol{v}_{s,j^{*}}^{(\ell)T} \,\, \boldsymbol{0}^{T}\big] \boldsymbol{v}^{(\ell)}_{k}\big|
\end{align}
where, without loss of generality, a specific state convention is assumed such that position states are first in element order.
\subsection{Recursion and Role of Negative Information}
The splitting procedure is applied recursively, as detailed in Algorithm~\ref{alg:SplitForFov}.  The recursion is terminated when no remaining components satisfy the criteria for splitting.  Each recursion further refines the \ac{gm} near the \ac{fov} bounds to improve the approximations of~(\ref{eq:InSTimesPApproxGm})-(\ref{eq:NotInSTimesPApproxGm}).  However, because a Gaussian component's split approximation (\ref{eq:SplitMultivariate}) does not perfectly replicate the original component, a small error is induced with each split.  Given enough recursions, this error may become dominant.  In the authors' experience, the recursion is terminated well before the cumulative split approximation error dominates.

\begin{algorithm}
\begin{algorithmic}
  \caption{\texttt{split\_for\_fov(}$\{w^{(\ell)}, \mathbf{m}^{(\ell)}, \mathbf{P}^{(\ell)}\}_{\ell=1}^{L}$, $w_{\min}$, $\mathcal{S}$, $R$, $\lambda$\texttt{)}}
  \label{alg:SplitForFov}
  \STATE \texttt{split} $\leftarrow \{\}$, \texttt{no\_split} $\leftarrow \{\}$
  \IF {$L=0$}
    \RETURN \texttt{split}
  \ENDIF
  \FOR {$\ell=1,\dots,L$}
    \IF {$w^{(\ell)}<w_{\min}$}
      \STATE add $\{w^{(\ell)}, \mathbf{m}^{(\ell)}, \mathbf{P}^{(\ell)}\}$ to \texttt{no\_split}
      \STATE \textbf{continue}
    \ENDIF
    \STATE Compute $\mathcal{S}^{(\ell)}_{y}$ according\ to~(\ref{eq:Sz})
    \IF {$\varrho_{\mathcal{S}_{y}^{(\ell)}}=1$}
    \STATE add $\{w^{(\ell)}, \mathbf{m}^{(\ell)}, \mathbf{P}^{(\ell)}\}$ to \texttt{no\_split}
    \ELSE
    \STATE $j^{*} \gets $ Eq.~(\ref{eq:BestPositionSplitDirection}) , $k^{*} \gets $ Eq.~(\ref{eq:BestFullStateSplitDirection})
    \STATE $\{w^{(\ell,j)}, \mathbf{m}^{(\ell,j)}, \mathbf{P}^{(\ell,j)}\}_{j=1}^{R} \gets$ Eq.~(\ref{eq:SplitMultivariate}) with $k=k^{*}$
    \STATE add $\{w^{(\ell,j)}, \mathbf{m}^{(\ell,j)}, \mathbf{P}^{(\ell,j)}\}_{j=1}^{R}$ to \texttt{split} \ENDIF
  \ENDFOR
  \STATE \texttt{split}$\leftarrow$\texttt{split\_for\_fov(split}, $w_{\min}$, $\mathcal{S}$, $R$, $\lambda${)}
  \RETURN \texttt{split} $\cup$ \texttt{no\_split}
\end{algorithmic}
\end{algorithm}

One of the many potential applications of the recursive algorithm presented in this section involves incorporating the evidence of non-detections, or negative information, in single- or multi-object filtering. To demonstrate, a single-object filtering problem with a bounded square \ac{fov} is considered where, in three subsequent sensor reports, no object is detected. The true object position and constant velocity are unknown but are distributed according to a known \ac{gm} \ac{pdf} at the first time step. As the initial \ac{pdf} is propagated over time, the position-marginal \ac{pdf} travels from left to right, as pictured in Figure~\ref{fig:GaussSplitRectFov}. For simplicity, the probability of detection inside the \ac{fov} is assumed equal to one. At each time step, the \ac{gm} is refined by Algorithm \ref{alg:SplitForFov} using $w_{\min}=0.01$, $R=3$, and $\lambda=0.001$. Then, the mean-based partition approximation~(\ref{eq:NotInSTimesPApproxGm}) is applied and the updated filtering density~(\ref{eq:1m1sp}) is found.
The results shown in Figure~\ref{fig:GaussSplitRectFov} are obtained using a \textsc{Matlab} implementation of Algorithm~\ref{alg:SplitForFov}.
  When executed on an Apple M1 Ultra processor with $64$~GB RAM, the total execution time (over three time steps) of Algorithm~\ref{alg:SplitForFov} is $0.176$ seconds, which translates to $<60$ milliseconds per time step.
As in many \ac{gm}-based filters, the number of components may increase over time but can be reduced as needed through component merging and pruning.

\newlength\gausssplitwidth
\setlength\gausssplitwidth{2.85cm}
\begin{figure}[htbp]
  \centering
  \includegraphics[width=\gausssplitwidth]{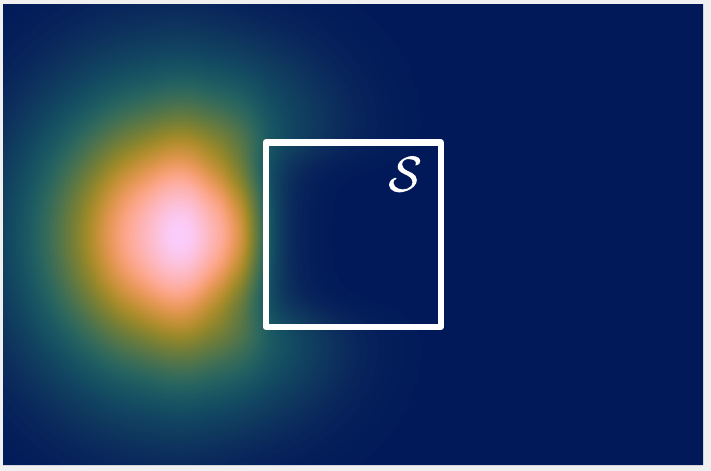}
  \includegraphics[width=\gausssplitwidth]{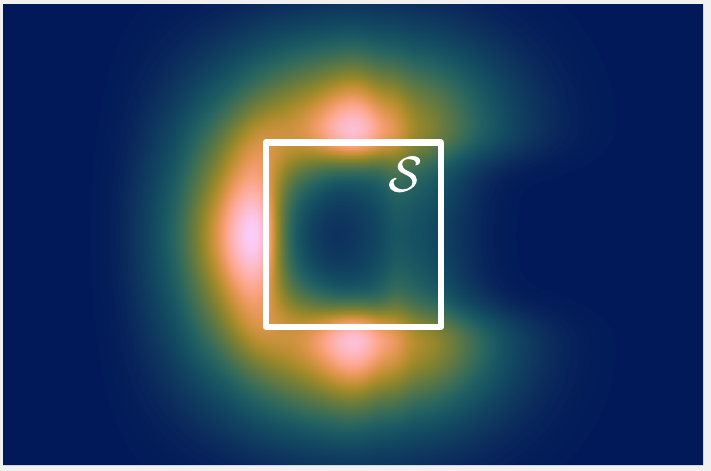}
  \includegraphics[width=\gausssplitwidth]{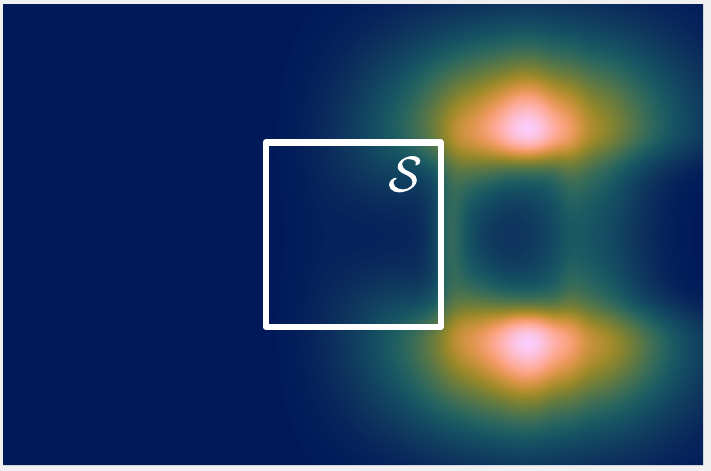}
  \caption{Negative information, comprised of absence of detections inside the sensor FoV $\mathcal{S}$, is used to update the object pdf as the object moves across the ROI.}
  \label{fig:GaussSplitRectFov}
\end{figure}

\subsection{Splitting for Multiple Regions}
\label{sec:SplittingMultipleFieldsOfView}
The presented splitting approach can be extended to accommodate multiple closed subsets, which may represent the \acp{fov} in a multi-sensor network or imprecise measurements that take the form of multiple closed subsets, as is shown in Section~\ref{sec:application_to_imprecise_measurements}.
For ease of exposition, the multi-region method is developed in the context of multiple \acp{fov} with the awareness that the regions can be any bounded sets.
Consider the case where the \ac{gm} is to be partitioned about the boundaries of $N_{\mathrm{s}}$ \acp{fov} $\{\mathcal{S}^{(\imath)}\}_{\imath=1}^{N_{\mathrm{s}}}$.
One simple approach to incorporate the multiple \acp{fov} is to recursively apply Algorithm~\ref{alg:SplitForFov} for each \ac{fov}.
Recall from Section~\ref{sec:position_split_direction}, however, that the direction order in which components are split ultimately determines the total number of components generated.
Thus, by the described naive approach, the resulting mixture size inherently depends on the order by which the \acp{fov} are processed, which is undesirable.

Instead, the remainder of this subsection establishes a multi-\ac{fov} splitting algorithm that is invariant to \ac{fov} order.
Given $\mathcal{S}^{(\imath)}$, denote by $\mathcal{S}_{y}^{(\imath,\ell)}$ the resulting transformed \ac{fov} for component $\ell$ via application of~(\ref{eq:Sz}).
Then, an inclusion variable similar to~(\ref{eq:ComponentInclusionVariable}) is established as
\begin{align}
  \label{eq:ComponentInclusionVariableMultiSensor}
  d_{i_{1},\dots, i_{n_{s}}}^{(\imath, \ell)}
  \triangleq
  1_{\mathcal{S}_{y}^{(\imath,\ell)}}(\bar{\mathbf{y}}_{i_{1}, \dots, i_{n_{s}}})
\end{align}
In each transformed \ac{fov}, grid points are either totally excluded or totally included if and only if
\begin{align}
  \varrho_{\{\mathcal{S}_{y}\}}^{{(\ell)}}
  =
  \prod_{\imath=1}^{N_{\mathrm{s}}}
  \prod_{G} {\delta_{d_{i_{1}', \dots, i_{n_{s}}'}^{(\imath, \ell)}}}(d_{i_{1},\dots,i_{n_{s}}}^{(\imath,\ell)})
\end{align}
is equal to unity, which indicates that a component does not require splitting.
If a component is to be split, the direction is chosen to minimize the ultimate mixture size, as discussed in Section~\ref{sec:position_split_direction}.
This is accomplished by identifying grid planes that are either consistently included/excluded in each \ac{fov}.
Consistency of the plane of constant $y_{j}=\bar{y}_{j}(l)$ is indicated by
\begin{align}
  \label{eq:ConsistentPlaneIndicatorMultiFov}
  s_{j}^{(\ell)}(l) =
  \prod_{\imath=1}^{N_{\mathrm{s}}}
\prod_{G,i_{j}=l} {\delta_{d_{i_{1}',\dots, i_{j}, \dots, i_{n_{s}}'}^{(\imath, \ell)}}}
  (d_{i_{1},\dots, i_{j}, \dots, i_{n_{s}}}^{(\imath,\ell)})
\end{align}
equal to unity.
By this multi-\ac{fov} generalized indicator function, the optimal position split direction is found via~(\ref{eq:BestPositionSplitDirection}).
The complete multi-\ac{fov} splitting algorithm is summarized in Algorithm~\ref{alg:SplitForMultiFov}.

\begin{algorithm}
\begin{algorithmic}
  \caption{\texttt{split\_for\_multifov(}$\{w^{(\ell)}, \mathbf{m}^{(\ell)}, \mathbf{P}^{(\ell)}\}_{\ell=1}^{L}$, $w_{\min}$, $\{\mathcal{S}^{(\imath)}\}_{\imath=1}^{N_{\mathrm{s}}}$, $R$, $\lambda$\texttt{)}}
  \label{alg:SplitForMultiFov}
  \STATE \texttt{split} $\leftarrow \{\}$, \texttt{no\_split} $\leftarrow \{\}$
  \IF {$L=0$}
    \RETURN \texttt{split}
  \ENDIF
  \FOR {$\ell=1,\dots,L$}
    \IF {$w^{(\ell)}<w_{\min}$}
      \STATE add $\{w^{(\ell)}, \mathbf{m}^{(\ell)}, \mathbf{P}^{(\ell)}\}$ to \texttt{no\_split}
      \STATE \textbf{continue}
    \ENDIF
    \FOR {$\imath=1,\hdots,N_{\mathrm{s}}$}
      \STATE compute $\mathcal{S}^{(\imath, \ell)}_{y}$ according to Eq.~(\ref{eq:Sz})
    \ENDFOR
    \IF {$\varrho_{\{\mathcal{S}_{y}\}}^{(\ell)}=1$}
    \STATE add $\{w^{(\ell)}, \mathbf{m}^{(\ell)}, \mathbf{P}^{(\ell)}\}$ to \texttt{no\_split}
    \ELSE
    \STATE $j^{*} \gets $ Eq.~(\ref{eq:BestPositionSplitDirection}) , $k^{*} \gets $ Eq.~(\ref{eq:BestFullStateSplitDirection})
    \STATE $\{w^{(\ell,j)}, \mathbf{m}^{(\ell,j)}, \mathbf{P}^{(\ell,j)}\}_{j=1}^{R} \gets$ Eq.~(\ref{eq:SplitMultivariate}) with $k=k^{*}$
    \STATE add $\{w^{(\ell,j)}, \mathbf{m}^{(\ell,j)}, \mathbf{P}^{(\ell,j)}\}_{j=1}^{R}$ to \texttt{split} \ENDIF
  \ENDFOR
  \STATE \texttt{split}$\leftarrow$\texttt{split\_for\_multifov(split}, $w_{\min}$, $\{\mathcal{S}^{(\imath)}\}_{\imath=1}^{N_{\mathrm{s}}}$, $R$, $\lambda${)}
  \RETURN \texttt{split} $\cup$ \texttt{no\_split}
\end{algorithmic}
\end{algorithm}

The set inputs $\{\mathcal{S}^{(\imath)}\}$ in Algorithm~\ref{alg:SplitForMultiFov} are not restricted to \acp{fov} and can represent any regions.
For example, two regions relevant to the human-robot interaction depicted in Fig.~\ref{fig:MugOnTable} are the human observer's binocular \ac{fov} and the tabletop region.
The application of Algorithm~\ref{alg:SplitForMultiFov} with respect to these two regions then enables the incorporation of the observation ``The mug is on the table'' in a \ac{gm} Bayes filter, as is discussed in the following section.
\section{Application to Imprecise Measurements}%
\label{sec:application_to_imprecise_measurements}
This section presents the application of the splitting algorithm to estimation problems involving imprecise measurements.
Unlike traditional vector-type measurements, imprecise measurements are non-specific, yet still contain valuable information.
Examples of imprecise measurements include natural language statements \cite{BishopFusionNaturalLanguageRandomSet13, RisticBernoulliFilterTutorial13}, inference rules \cite[Sec.\ 22.2.4]{MahlerAdvancesStatisticalMultitargetFusion14}, and received signal strength type measurements under path-loss uncertainty \cite{RisticBernoulliFilterTutorial13, PatwariLocatingTheNodesWirelessSensorNetwork05}.
This section demonstrates the estimation of a person's location and velocity as they move through a public space using imprecise natural language measurements, as originally posed in \cite{RisticBernoulliFilterTutorial13}.
Tracking is performed using a new \ac{gm} Bernoulli filter for imprecise measurements, as discussed in the following subsections.

\subsection{Imprecise Measurements}%
\label{sec:imprecise_imprecise_measurements}
Imprecise measurements, such as those from natural language statements, can be modeled as \acp{rfs} and specified using \emph{generalized likelihood functions}.
For example, the statement
\begin{equation}
  S = \textrm{``Felice is near the taco stand''}
\end{equation}
provides some evidence about Felices's location, yet is not mutually exclusive\footnote{In fact, this statement can further be considered vague or fuzzy due to uncertainty in the observer's definition of ``near'' \cite[p.\ 266]{GoodmanMathematicsOfDataFusion97}.} \cite[p.\ 104, 126]{MahlerStatisticalMultitargetFusion07}.
For simplicity, this paper adopts from \cite[p.\ 105]{MahlerStatisticalMultitargetFusion07} the definition of being ``near'' a point $\mathbf{z}_{0}$ as belonging to a disc $\boldsymbol{\zeta} \subset \mathbb{Z}$ of radius $D$:
\begin{equation}
  \boldsymbol{\zeta} = \{\mathbf{z}: \|\mathbf{z}-\mathbf{z}_{0}\| \leq D\}
\end{equation}
Although this specific natural language statement interpretation is considered for simplicity, the presented approach does not preclude more sophisticated models, such as in \cite{BishopFusionNaturalLanguageRandomSet13,TseMixtureOfStatements18}.
The associated generalized likelihood function for this imprecise measurement is
\begin{align}
  \label{eq:GeneralizedLikelihood}
  \tilde{g}(\boldsymbol{\zeta}|\mathbf{x}) = P\{\mathbf{z} \in \boldsymbol{\zeta}\}
  = P\{ \mathbf{h}(\mathbf{x}) \in \boldsymbol{\zeta}\}
\end{align}
where $\mathbf{h} \, : \mathbb{X} \mapsto \mathbb{Z}$ is the deterministic mapping from the state space to the measurement space \cite{RisticBernoulliFilterTutorial13}.
Generalized likelihood functions, such as those for natural language statements, are often nonlinear in $\mathbf{x}$.
Through the presented Gaussian splitting approach and expansion of the nonlinear likelihood function about the \ac{gm} component means, \ac{gm} \ac{rfs} filters can accommodate imprecise measurements, as demonstrated in the context of the \ac{rfs} Bernoulli filter in the following subsection.

\subsection{Bernoulli Filter for Imprecise Measurements}
\label{sec:BernoulliFilter}
The Bernoulli filter is the Bayes-optimal filter for tracking a single object in the presence of false alarms, misdetections, and unknown object birth/death \cite[Sec.\ 14]{MahlerStatisticalMultitargetFusion07}.
A Bernoulli distribution is parameterized by a probability of object existence $r$ and state \ac{pdf} $p(\mathbf{x})$.
The \ac{fisst} density of a Bernoulli \ac{rfs} is \cite[p.\ 516]{MahlerStatisticalMultitargetFusion07}
\begin{align}
  f(X) =
  \begin{cases}
    1 - r, & \textrm{if } X = \emptyset\\
    r \cdot p(\mathbf{x}), & \textrm{if } X = \{\mathbf{x}\}
  \end{cases}
\end{align}

Denote by $p_b$ the conditional probability that the object is born given that it did not exist in the previous time step.
Similarly, denote by $p_{S}$ the conditional probability that the object survives to the next time step.
The initial state of an object born at time $k$ is assumed to be distributed according to the birth spatial density $b_k(\mathbf{x})$.
Then, by the \ac{fisst} generalized Chapman-Kolmogorov equation, the Bernoulli filter prediction equations are \cite[p.\ 519]{MahlerStatisticalMultitargetFusion07}
\begin{align}
  \label{eq:BernoulliPredictedSpatialDensity}
  p_{k|k-1}(\mathbf{x}) &=
  \frac{p_b \cdot (1 - r_{k-1|k-1}) b_{k|k-1}(\mathbf{x})}{r_{k|k-1}} \\
  &\qquad+ \frac{p_{S} r_{k-1|k-1} \int \pi_{k|k-1}(\mathbf{x}|\mathbf{x}') p_{k-1|k-1}(\mathbf{x}')\mathrm{d} \mathbf{x}'}{r_{k|k-1}}\nonumber\\
  \label{eq:BernoulliPredictedExistence}
  r_{k|k-1} &= p_{b} \cdot (1-r_{k-1| k-1}) + p_{S} r_{k-1|k-1}
\end{align}
where $\pi_{k|k-1}(\mathbf{x}|\mathbf{x}')$ is the single-object state transition density.
Suppose that the spatial density and birth density are \acp{gm} and that the transition is linear-Gaussian:
\begin{align}
  p_{k-1|k-1}(\mathbf{x}) &=
  \sum_{\ell=1}^{L_{k-1}}
  w_{k-1}^{(\ell)}
  \mathcal{N}(\mathbf{x}; \, \mathbf{m}_{k-1}^{(\ell)}, \, \mathbf{P}_{k-1}^{(\ell)})\\
  b_{k|k-1}(\mathbf{x}) &=
  \sum_{\ell=1}^{L_{b,k}} \hat{w}_{b,k}^{(\ell)}
  \mathcal{N}(\mathbf{x}; \, \mathbf{m}_{b,k}^{(\ell)}, \, \mathbf{P}_{b,k}^{(\ell)})\\
  \pi_{k|k-1}(\mathbf{x}|\mathbf{x}') &=
  \mathcal{N}(\mathbf{x}; \, \mathbf{F}_{k-1}\mathbf{x}', \, \mathbf{Q}_{k-1})
\end{align}
Then, the predicted spatial density at $k$ is the sum of two \acp{gm}, given as
\begin{align}
  \label{eq:GmBernoulliPredictedSpatialDensityTwoGms}
  p_{k|k-1}(\mathbf{x}) &=
  \sum_{\ell=1}^{L_{b,k}}
  w_{b,k}^{(\ell)}
  \mathcal{N}(\mathbf{x}; \, \mathbf{m}_{b,k}^{(\ell)}, \, \mathbf{P}_{b,k}^{(\ell)})\\
  &+
  \sum_{\ell=1}^{L_{k-1}}
  w_{S,k|k-1}^{(\ell)}
  \mathcal{N}(\mathbf{x}; \, \mathbf{m}_{S,k|k-1}^{(\ell)}, \, \mathbf{P}_{S,k|k-1}^{(\ell)})
  \nonumber
\end{align}
where
\begin{align}
  \label{eq:GmBernoulliBirthWeightsUpdated}
  w_{b,k}^{(\ell)} &= \hat{w}_{b,k}^{(\ell)}
  \frac{p_b \cdot (1 - r_{k-1|k-1})}{r_{k|k-1}} \\
  \label{eq:GmBernoulliPredictedWeight}
  w_{S,k|k-1}^{(\ell)} &=
  w_{k-1}^{(\ell)} \frac{p_{S} r_{k-1|k-1}}{r_{k|k-1}}\\
  \label{eq:GmBernoulliPredictedMean}
  \mathbf{m}_{S,k|k-1}^{(\ell)} &=
  \mathbf{F}_{k-1} \mathbf{m}_{k-1}^{(\ell)}\\
  \label{eq:GmBernoulliPredictedCov}
  \mathbf{P}_{S,k|k-1}^{(\ell)} &=
  \mathbf{F}_{k-1} \mathbf{P}_{k-1}^{(\ell)} \mathbf{F}_{k-1}^{T} + \mathbf{Q}_{k-1}
\end{align}
The predicted spatial density (\ref{eq:GmBernoulliPredictedSpatialDensityTwoGms}) can thus be expressed as a combined \ac{gm} of the form
\begin{align}
  p_{k|k-1}(\mathbf{x}) &=
  \sum_{\ell=1}^{L_{k|k-1}}
  w_{k|k-1}^{(\ell)}
  \mathcal{N}(\mathbf{x}; \, \mathbf{m}_{k|k-1}^{(\ell)}, \, \mathbf{P}_{k|k-1}^{(\ell)})
\end{align}
where $\sum_{\ell=1}^{L_{k|k-1}} w_{k|k-1}^{(\ell)}=1$.

The \ac{fov}-dependent probability of detection function is given by
\begin{align}
  \label{eq:ProbOfDetectFoVDependent}
  p_{D}(\mathbf{x}; \mathcal{S}_{k})
  =
  1_{\mathcal{S}_{k}}(\mathbf{x}) p_{D}(\mathbf{s})
\end{align}
where the single-argument function $p_{D}(\mathbf{s})$ is the corresponding probability of detection for an unbounded \ac{fov}.
The measurement $\Upsilon_{k}$ is then a finite set
\begin{align}
  \Upsilon_{k} = \{\boldsymbol{\zeta}_{1},\, \hdots, \, \boldsymbol{\zeta}_{m_{k}}\}
  \in \mathcal{F}(\mathfrak{Z})
\end{align}
comprised of false alarms and a potentially empty imprecise measurement due a true object, where $\mathfrak{Z}$ is the set of all closed subsets of $\mathbb{Z}$ and $\mathcal{F}(\mathfrak{Z})$ is the space of all finite subsets of $\mathfrak{Z}$, as shown in \cite[Ch.\ 5]{MahlerStatisticalMultitargetFusion07}.
Assume that false alarms are Poisson distributed with rate $\lambda_{c}$ and spatial density $\tilde{c}(\boldsymbol{\zeta})$.
Then, the posterior state density and probability of existence are given by
\begin{align}
  \label{eq:BernoulliStatePdfUpdate}
  p_{k|k}(\mathbf{x}) &= \frac{1-p_{D}(\mathbf{x}; \mathcal{S}_{k}) + p_{D}(\mathbf{x}; \mathcal{S}_{k}) \sum\limits_{\boldsymbol{\zeta} \in \Upsilon_{k}} \frac{\tilde{g}_{k}(\boldsymbol{\zeta} | \mathbf{x})}{\lambda_{c} \tilde{c}(\boldsymbol{\zeta})}}{1 - \Delta_k} p_{k|k-1}(\mathbf{x}) \\
  \label{eq:BernoulliProbExistenceUpdate}
  r_{k|k} &= \frac{1- \Delta_k}{1 - r_{k|k-1}\Delta_k} r_{k|k-1}
\end{align}
where
\begin{align}
  \Delta_{k} =& \int p_{D}(\mathbf{x}; \mathcal{S}_{k}) p_{k|k-1}(\mathbf{x}) \mathrm{d} \mathbf{x} \nonumber \\
              &- \sum_{\boldsymbol{\zeta}\in \Upsilon_{k}}
              \frac{\int p_{D}(\mathbf{x}; \mathcal{S}_{k}) \tilde{g}_{k}(\boldsymbol{\zeta}|\mathbf{x}) p_{k|k-1}(\mathbf{x}) \mathrm{d} \mathbf{x}}{\lambda_{c} \tilde{c}(\boldsymbol{\zeta})}
\end{align}
which is a generalization of the result shown in \cite{GningBernoulliParticleFilterTripleUncertainty12} for state-dependent probability of detection.%

Because~(\ref{eq:BernoulliStatePdfUpdate}) involves products of indicator functions and \acp{gm}, the resulting posterior density will not be a \ac{gm} in general.
Instead, the state-dependent probability of detection and generalized likelihood function can be expanded about the \ac{gm} component means (see Appendix~\ref{a:change1_taylor_series_expansion_about_means}), giving
\begin{align}
  \label{eq:GmBernoulliUpdatedMixture}
  p_{k|k}(\mathbf{x})
  &=
  \sum_{\ell=1}^{L_{k|k}}
  w_{k|k}^{(\ell)}
  \mathcal{N}(\mathbf{x};\, \mathbf{m}_{k|k}^{(\ell)},\, \mathbf{P}_{k|k}^{(\ell)}) \\
  \label{eq:GmBernoulliUpdatedWeight}
  w_{k|k}^{(\ell)} &=
  \frac{w_{k|k-1}^{(\ell)}}{1 - \Delta_{k}}
  \Bigg(
    1 - p_{D}(\mathbf{m}_{k|k-1}^{(\ell)}; \mathcal{S}_{k})
    \nonumber \\
   &\quad
    + p_{D}(\mathbf{m}_{k|k-1}^{(\ell)}; \mathcal{S}_{k})
    \sum_{\boldsymbol{\zeta} \in \Upsilon_{k}}
    \frac{\tilde{g}_{k}(\boldsymbol{\zeta} | \mathbf{m}_{k|k-1}^{(\ell)})}{\lambda_{c}\tilde{c}(\boldsymbol{\zeta})}
  \Bigg)\\
  \label{eq:GmBernoulliUpdateDelta}
  \Delta_{k}
  &=
  \sum_{\ell=1}^{L_{k|k-1}}
  w_{k|k-1}^{(\ell)} p_{D}(\mathbf{m}_{k|k-1}^{(\ell)}; \mathcal{S}_{k}) \\
  &-
  \sum\limits_{\boldsymbol{\zeta} \in \Upsilon_{k}}
  \frac{
    \sum\limits_{\ell=1}^{L_{k|k-1}} w_{k|k-1}^{(\ell)} p_{D}(\mathbf{m}_{k|k-1}^{(\ell)}; \mathcal{S}_{k})
    \tilde{g}_{k}(\boldsymbol{\zeta}|\mathbf{m}_{k|k-1}^{(\ell)})
        }{
        \lambda_{c} \tilde{c}(\boldsymbol{\zeta})
  }\nonumber\\
  \label{eq:GmBernoulliUpdatedMean}
  \mathbf{m}_{k|k}^{(\ell)} &= \mathbf{m}_{k|k-1}^{(\ell)} \\
  \label{eq:GmBernoulliUpdatedCov}
  \mathbf{P}_{k|k}^{(\ell)} &= \mathbf{P}_{k|k-1}^{(\ell)}
\end{align}
The approximation error due to the zeroth-order expansion in~(\ref{eq:GmBernoulliUpdatedWeight}) and~(\ref{eq:GmBernoulliUpdateDelta}) depends on the \ac{gm} resolution near points of strong nonlinearity.
In a high resolution mixture containing many components with small covariance matrices, the region about each mean in which the local approximation must be valid is correspondingly smaller compared to a low resolution mixture \cite{AlspachNonlinearGaussianSums72}.
Therefore, the recursive splitting method is employed to refine the mixture in nonlinear regions--specifically around $\partial\mathcal{S}_{k}$ and $\partial \boldsymbol{\zeta}_{(\cdot)}$--before computing the posterior \ac{gm}~(\ref{eq:GmBernoulliUpdatedMixture}).
Then, the resulting posterior \ac{gm} is reduced using one of many available algorithms for \ac{gm} reduction \cite{WestApproximatingPosteriorMixtures93, RunnallsKullbackLeiblerGaussianMixtureReduction07, SalmondMixtureReduction09, CrouseGaussianMixtureReduction11}.
This process, referred to as the \ac{gm} Bernoulli filter for imprecise measurements, is summarized in Algorithm~\ref{alg:GmBernoulliImprecise}.
\begin{algorithm*}
\begin{algorithmic}
  \footnotesize
\caption{GM Bernoulli Filter for Imprecise Measurements}
  \label{alg:GmBernoulliImprecise}
  \STATE \textbf{given}  $r_{0|0}, p_{0|0}(\mathbf{x})$
  \FOR {$k=1,\dots,K$}
    \STATE Compute $r_{k|k-1}$ according to (\ref{eq:BernoulliPredictedExistence})
    \STATE Compute $\{w_{S,k|k-1}^{(\ell)},\mathbf{m}_{S,k|k-1}^{(\ell)}, \mathbf{P}_{S,k|k-1}^{(\ell)}\}_{\ell=1}^{L_{k|k-1}}$ according to (\ref{eq:GmBernoulliPredictedWeight})-(\ref{eq:GmBernoulliPredictedCov})
    \STATE Compute $\{w_{b,k}^{(\ell)}\}_{\ell=1}^{L_{b,k}}$ according to (\ref{eq:GmBernoulliBirthWeightsUpdated})
    \STATE $\{w_{k|k-1}^{(\ell)}, \mathbf{m}_{k|k-1}^{(\ell)}, \mathbf{P}_{k|k-1}^{(\ell)}\}_{\ell=1}^{L_{k|k-1}}
    \gets
    \{w_{S,k|k-1}^{(\ell)},\mathbf{m}_{S,k|k-1}^{(\ell)}, \mathbf{P}_{S,k|k-1}^{(\ell)}\}_{\ell=1}^{L_{k-1}} \cup \{w_{b,k|k-1}^{(\ell)},\mathbf{m}_{b,k|k-1}^{(\ell)}, \mathbf{P}_{b,k|k-1}^{(\ell)}\}_{\ell=1}^{L_{b,k}}$%
    \STATE $\{w_{k|k-1}^{(\ell)}, \mathbf{m}_{k|k-1}^{(\ell)}, \mathbf{P}_{k|k-1}^{(\ell)}\}_{\ell=1}^{L_{k|k-1}}$%
    $\gets$%
    \texttt{split\_for\_multifov(}$\{w_{k|k-1}^{(\ell)}, \mathbf{m}_{k|k-1}^{(\ell)}, \mathbf{P}_{k|k-1}^{(\ell)}\}_{\ell=1}^{L_{k|k-1}}$, $w_{\min}$, $\{\mathcal{S}_{k}\}\cup \Upsilon_{k}$, $R$, $\lambda$\texttt{)}
    \STATE Compute $\Delta_{k}$ according to (\ref{eq:GmBernoulliUpdateDelta})
    \STATE Compute $r_{k|k}$ according to (\ref{eq:BernoulliProbExistenceUpdate})
    \STATE Compute $\{w_{k|k}^{(\ell)},\mathbf{m}_{k|k}^{(\ell)}, \mathbf{P}_{k|k}^{(\ell)}\}_{\ell=1}^{L_{k|k}}$ according to (\ref{eq:GmBernoulliUpdatedWeight}),(\ref{eq:GmBernoulliUpdatedMean}),(\ref{eq:GmBernoulliUpdatedCov})
    \STATE $\{w_{k|k}^{(\ell)},\mathbf{m}_{k|k}^{(\ell)}, \mathbf{P}_{k|k}^{(\ell)}\}_{\ell=1}^{L_{k|k}}$ $\gets$ \texttt{reduce(} $\{w_{k|k}^{(\ell)},\mathbf{m}_{k|k}^{(\ell)}, \mathbf{P}_{k|k}^{(\ell)}\}_{\ell=1}^{L_{k|k}}$\texttt{)}
  \ENDFOR
\end{algorithmic}
\end{algorithm*}
\subsection{Airport Tracking Example}%
\label{sec:airport_tracking_example}
The recursive splitting approach is demonstrated in the context of tracking a person of interest through a crowded airport.
This problem was originally posed in \cite{RisticBernoulliFilterTutorial13} and solved using a \ac{pf} implementation of the Bernoulli filter.
The object state is defined as
\begin{align}
\mathbf{x}_{k}^T = [x_{k} \quad y_{k} \quad \dot{x}_{k} \quad \dot{y}_{k}] = [\mathbf{s}_{k}^T \quad \mathbf{v}_{k}^{T}]
\end{align}
where dimensionless distance units are used throughout.
Measurements of the object are composed of natural language statements describing the person's current location in the form $Z_k = \{\zeta_{k,1}, \hdots, \zeta_{k, m_{k}}\}$, where $m_{k}$ is the number of statements received at time $k$ and
\begin{align}
  \label{eq:NlsTargetIsNearTheAnchor}
  \zeta = a \implies \textrm{the object is near the anchor } a
\end{align}
In (\ref{eq:NlsTargetIsNearTheAnchor}), the integer $a\in \mathbb{A}\subset \mathbb{N}$ represents a fixed anchor, such as a taco stand or coffee shop, with corresponding known position $\mathbf{r}_{a} \in \mathbb{Z}$.
Observers sometimes report incorrect statements (as false alarms) and sometimes fail to report true statements (as misdetections).
The corresponding generalized likelihood function is
\begin{align}
  \label{eq:NlsGeneralizedLikelihood}
  \tilde{g}_k(\zeta =a \, | \, \mathbf{x}_k)
  =
  \begin{cases}
    1 & \textrm{if } \|\mathbf{s}_k - \mathbf{r}_a \| \leq 2 d_a /3 \\
    0 & \textrm{otherwise}
  \end{cases}
\end{align}
where $d_a$ is the distance between anchor $a$ and its nearest neighboring anchor.
If the object is within $2d_a/3$ of anchor $a$, the natural language statement reports that the object is near $a$ (unless misdetected).
Defining the compact subset
\begin{align}
  \mathcal{A}_{a} = \{\mathbf{s} \, : \, \|\mathbf{s} - \mathbf{r}_{a}\| \leq 2d_{a}/3 \} \,,
\end{align}
the generalized likelihood function (\ref{eq:NlsGeneralizedLikelihood}) can be written in terms of an indicator function as
\begin{align}
  \tilde{g}_k(\zeta = a \, | \, \mathbf{x}_k) = 1_{\mathcal{A}_{a}}(\mathbf{s}_{k})
\end{align}
By this likelihood function, (\ref{eq:GmBernoulliUpdatedWeight})-(\ref{eq:GmBernoulliUpdateDelta}) simplify to
\begin{align}
  \label{eq:GmBernoulliUpdatedSimplifiedWeight}
  w_{k|k}^{(\ell)} &=
  \frac{w_{k|k-1}^{(\ell)}}{1 - \Delta_{k}}
  \Bigg(
    1 - p_{D}(\mathbf{m}_{k|k-1}^{(\ell)}; \mathcal{S}_{k})
    \nonumber \\
   &\quad
    + p_{D}(\mathbf{m}_{k|k-1}^{(\ell)}; \mathcal{S}_{k})
    \sum_{\zeta \in Z_{k}}
    \frac{1_{\mathcal{A}_{\zeta}}(\mathbf{m}_{s,k|k-1}^{(\ell)})}{\lambda_{c}\tilde{c}(\zeta)}
  \Bigg)\\
  \label{eq:GmBernoulliUpdateSimplifiedDelta}
  \Delta_{k}
  &=
  \sum_{\ell=1}^{L_{k|k-1}}
  w_{k|k-1}^{(\ell)} p_{D}(\mathbf{m}_{k|k-1}^{(\ell)}; \mathcal{S}_{k}) \\
  &-
  \sum_{\zeta \in Z_{k}}
  \frac{
    \sum\limits_{\ell=1}^{L_{k|k-1}} w_{k|k-1}^{(\ell)} p_{D}(\mathbf{m}_{k|k-1}^{(\ell)}; \mathcal{S}_{k})
    1_{\mathcal{A}_{\zeta}}(\mathbf{m}_{s,k|k-1}^{(\ell)}; \mathcal{S}_{k})
  }{
    \lambda_{c} \tilde{c}(\zeta)
  }\nonumber
\end{align}
where $\lambda_{c}$ denotes the clutter cardinality mean and the density of clutter $\tilde{c}(\zeta)$ is taken to be uniform over support $\mathbb{A}$.

\begin{figure}[htpb]
  \centering
  \ifx\undefined\usetikz
    \includegraphics{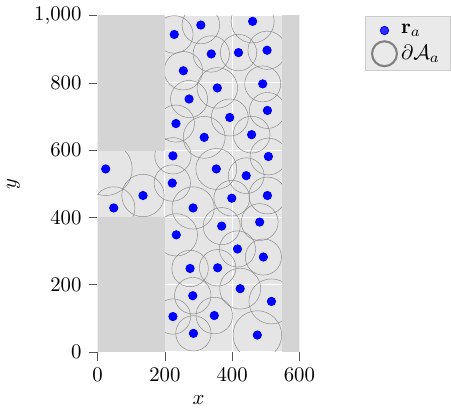}
  \else
    \input{figures/anchors.tex}%
  \fi
  \caption{Anchor locations and association extents.}%
  \label{fig:Anchors}
\end{figure}

The anchor locations and bounds $\partial \mathcal{A}_{a}$ are shown in Fig.~\ref{fig:Anchors}.
The gray shaded regions indicate exclusion regions the person cannot occupy due to physical barriers, and thus, $p_{k}(\mathbf{x})=0$ in these regions.
Detections are reported every $T_k=15\, [\textrm{s}]$ and include an average of $\lambda_{c}=0.25$ false detections.
True detections are reported with a probability of detection $p_{D}(\mathbf{x}_{k}; \mathcal{S}_{k})$ given by~(\ref{eq:ProbOfDetectFoVDependent}) with $ p_{D}(\mathbf{s}_{k}) = 0.9 $ and composite detection \ac{fov}
\begin{align}
  \mathcal{S}_{k} = \bigcup\limits_{a\in \mathbb{A}}\mathcal{A}_{a}
\end{align}
The object state is governed by the transition density
\begin{align}
  \pi_{k|k-1}(\mathbf{x}| \mathbf{x}') = \mathcal{N}(\mathbf{x}; \, \mathbf{F}_{k-1} \mathbf{x}', \, \mathbf{Q}_{k-1})
\end{align}
where
\begin{align}
  \mathbf{F}_k &=
  \begin{bmatrix}
    1 & 0 & T_k & 0\\
    0 & 1 & 0 & T_k\\
    0 & 0 & 1 & 0 \\
    0 & 0 & 0 & 1
  \end{bmatrix}\\
  \mathbf{Q}_{k} &=
  \begin{bmatrix}
    \frac{\varpi T_{k}^{3}}{3} & 0 & \frac{\varpi T_{k}^{2}}{2} & 0 \\
    0 & \frac{\varpi T_{k}^{3}}{3} & 0  & \frac{\varpi T_{k}^{2}}{2}\\
    \frac{\varpi T_{k}^{2}}{2} & 0 & \varpi T_{k} & 0 \\
    0 & \frac{\varpi T_{k}^{2}}{2} & 0 & \varpi T_{k}
  \end{bmatrix}
\end{align}
and $\varpi=0.004$ is the intensity of the process noise.

The simulated reports are processed by the \ac{gm} Bernoulli for imprecise measurements (Alg.~\ref{alg:GmBernoulliImprecise}) and the Bernoulli \ac{pf} \cite{RisticBernoulliFilterTutorial13} at each time step to obtain the posterior probability of existence and state density.

  The Bernoulli \ac{pf} is implemented using $5,000$ particles and a \ac{mcmc} move step to improve sample diversity, as described in \cite{RisticBernoulliFilterTutorial13}.

By splitting the density about the relevant anchor boundaries, the imprecise measurements are incorporated to refine the probabilistic belief and estimate the person's trajectory over time.

\begin{figure}[htpb]
  \centering
  \subfloat{%
      \ifx\undefined\usetikz
      \includegraphics{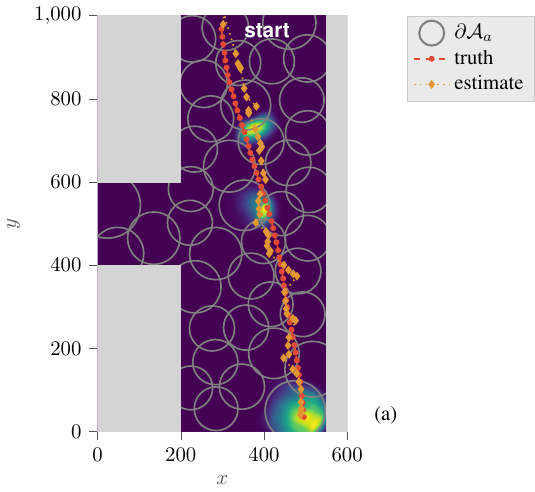}
      \else
      \input{figures/densities_over_time.tex}
      \fi
      \label{State:Estimates}
     }
     \hfill
  \subfloat{%
      \ifx\undefined\usetikz
      \includegraphics{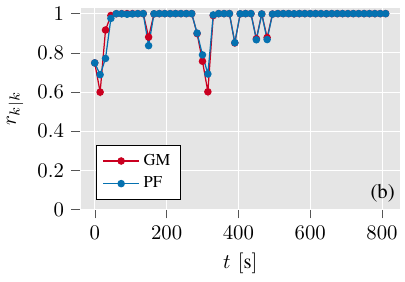}
      \else
      \input{figures/qkp.tex}
      \fi
      \label{ProbExist:Estimates}
     }
     \caption{(a) True trajectory and \ac{gm} Bernoulli filter state estimates over time, where position state densities are shown for time steps $k=15,25,55$ ($t=225,375,825\,[\textrm{s}]$) and (b)~posterior probability of existence over time.}%
  \label{fig:Estimates}
\end{figure}

The true trajectory, \ac{mmse} estimates, and densities at select time steps are shown in Fig.~\ref{State:Estimates}.
The Bernoulli \ac{pf} estimates and densities are omitted for clarity.
As shown, the true trajectory is consistently within the spatial distribution support.

The posterior probability of existence is shown over time in Fig.~\ref{ProbExist:Estimates}.
The probability of existence of the object is consistently near one, falling momentarily to $r_{k|k}=0.6$.
This drop in probability appropriately reflects the increased uncertainty after three consecutive misdetections (the latter two of which are due to the object traveling outside detection bounds).
As shown, the \ac{gm} and \ac{pf} approximations produce similar probability of existence estimates, where only slight differences are observed at times of non-detection.

The \ac{gm} Bernoulli filter for imprecise measurements is exceptionally computationally efficient, resulting in a total simulation time of $45.2$ seconds.
When applied to identical measurement data, the Bernoulli \ac{pf} simulation required $128.5$ seconds.
In fact, the largest computational bottleneck of the presented \ac{gm} approach is the \ac{gm} reduction step.
A two-pass reduction strategy was found to effectively balance computational cost and estimation accuracy.
The Mahalanobis distance-based merge strategy of \cite{VoGaussianMixturePhd06} quickly reduces the number of \ac{gm} components in the first pass.
Then, if needed, the \ac{kld}-based Runnals algorithm \cite{RunnallsKullbackLeiblerGaussianMixtureReduction07} further reduces the mixture size to $L_{\max}=100$.

\begin{figure}[htpb]
  \centering
  \subfloat{%
      \setlength{\figureheight}{5cm}
      \ifx\undefined\usetikz
      \includegraphics{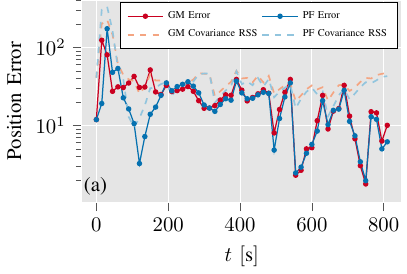}
      \else
      \input{figures/position_rss.tex}
      \fi
     }
     \hfill
     \subfloat{%
      \setlength{\figureheight}{5cm}
      \ifx\undefined\usetikz
      \includegraphics{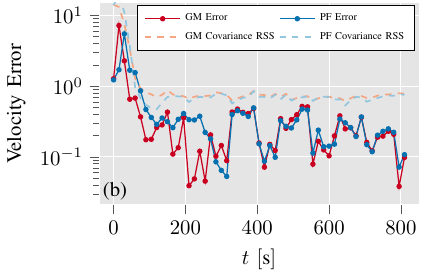}
      \else
      \input{figures/velocity_rss.tex}
      \fi
     }
     \caption{MMSE estimation error and conditional covariance RSS of position (a) and velocity (b) states.}%
  \label{fig:CovarianceRss}
\end{figure}

The state estimation performance is quantified using the \ac{mmse} estimate error and the \ac{rss} of the posterior conditional covariance and shown in Fig.~\ref{fig:CovarianceRss}.
The estimation performance of the \ac{gm} filter is very similar to the Bernoulli \ac{pf}, with neither method exhibiting a clear advantage in terms of estimation accuracy.
The velocity \ac{rss} quickly converges to a steady state of approximately $0.7 \, [\textrm{dist}/\textrm{s}]$, the lower bound of which is largely determined by the person's assumed maneuverability and associated process noise covariance.
Similarly, the largest uncertainty is observed near $k=21$ ($t=315 \, [\textrm{s}]$), after three consecutive misdetections.

While this example considers single-object estimation, the expansion approximation and splitting approach described in Section~\ref{sec:BernoulliFilter} is applicable to any \ac{gm} \ac{rfs} filter and thus can be used in multi-object estimation problems.
In the presented example of tracking a person of interest and its multi-object extension involving multiple persons of interest, the posterior \ac{rfs} density can be used to intelligently query or deploy resources to find or intercept persons of interest.
In this case, one particularly useful statistic is the probability that a given number of individuals are near a particular anchor.
This information is fully described by the \ac{rfs} \ac{fov} cardinality distribution, as presented in the following section.
\section{FoV Cardinality Distribution}
\label{sec:FovCardinality}
This section presents \acp{pmf} for the cardinality of objects inside a bounded \ac{fov} $\mathcal{S}$ given different global multi-object densities $f(\cdot)$.
Previous work derived expressions for the first and second moments of \ac{fov} cardinality distributions given Poisson, \ac{iidc} \cite{DelandeRegionalStatisticsPoissonIidc14}, and \ac{mb} \cite{DelandeRegionalVarianceMultiBernoulli14} global densities.
This section instead develops full \acp{pmf} expressions, from which first, second, or any higher-order moments can be easily obtained \cite[Ch.\ 30]{RileyMathematicalMethods06}.
A similar concept is discussed in \cite{MahlerAdvancesStatisticalMultitargetFusion14} in the context of ``censored'' \acp{rfs}, and a general expression is provided in terms of set derivatives and belief mass functions.
This paper presents a new direct approach to obtain \ac{fov} cardinality distributions based on conditional cardinality functions and derives new simplified expressions for representative \ac{rfs} distribution classes.
The Poisson, \ac{iidc}, \ac{mb}, and \ac{glmb} distributions are considered in Subsections~\ref{ss:poisson},~\ref{ss:iidc},~\ref{ss:mb}, and~\ref{ss:glmb}, respectively.

The probability of $n$ objects existing inside \ac{fov} $\mathcal{S}$ conditioned on $X$ can be written in terms of the indicator function as
\begin{align}
  \rho_{\mathcal{S}}(n \,|\, X) = \sum_{X^{n} \subseteq X} [1_{\mathcal{S}}(\cdot)]^{X^{n}} [1 - 1_{\mathcal{S}}(\cdot)]^{X \setminus X^{n}}
  \label{eq:ConditionalCardinality}
\end{align}
where the summation is taken over all subsets $X^{n}\subseteq X$ with cardinality $n$.  Given the \ac{rfs} density $f(X)$, the \ac{fov} cardinality distribution is obtained via the set integral as
\begin{align}
  \rho_{\mathcal{S}}(n) = \int \rho_{\mathcal{S}}(n\,|\,X) f(X) \delta X
\end{align}
Expanding the integral,
\begin{align}
  \label{eq:FovCardinalityGeneralExpanded}
  &\rho_{\mathcal{S}}(n) =  \\
  &\quad\sum_{m=n}^{\infty} \frac{1}{m!}
  \int\limits_{\mathbb{X}^{m}} \rho_{\mathcal{S}}(n\,|\,\{\mathbf{x}_{1} \varlist \mathbf{x}_{m}\}) f(\{\mathbf{x}_{1} \varlist \mathbf{x}_{m}\}) \mathrm{d} \mathbf{x}_{1} \prodvarlist \mathrm{d} \mathbf{x}_{m} \nonumber
\end{align}
\textit{Remark}: The results presented in this section can be trivially extended to express the predicted cardinality of object-originated \textit{detections} $Z$ (excluding false alarms) by noting that
\begin{align}
  \rho_{\mathcal{S}}(n_{Z}\,|\, X) = \sum_{X^{n} \subseteq X} [p_{D}(\cdot) 1_{\mathcal{S}}(\cdot)]^{X^{n}} [1 - p_{D}(\cdot)1_{\mathcal{S}}(\cdot)]^{X \setminus X^{n}}
\end{align}
where $n_{Z}=|Z|$.

\subsection{Poisson Distribution}
\label{ss:poisson}
The density of a Poisson-distributed \ac{rfs} is
\begin{align}
  f(X) = e^{-N_{X}} [D]^{X}
  \label{eq:PoissonDensity}
\end{align}
where $N_{X}$ is the global cardinality mean, and $D(\mathbf{x})$ is the \ac{phd}, or intensity function, of $X$, which is defined on the single-object space $\mathbb{X}$.  One important property of the \ac{phd} is that its integral over a closed set on $\mathbb{X}$ yields the expected number of objects within that set, i.e.
\begin{align}
  \label{eq:PoissonIntegralIsExpectedCardinality}
  E[|X\cap T|] = \int_{T} D(\mathbf{x}) \mathrm{d} \mathbf{x}
\end{align}
\begin{proposition}
  Given a Poisson-distributed \ac{rfs} with \ac{phd} $D(\mathbf{x})$ and global cardinality mean $N_{X}$, the cardinality of objects inside the \ac{fov} $\mathcal{S}\subseteq \mathbb{X}_{s}$ is distributed according to
  \begin{align}
    \rho_{\mathcal{S}}(n) = \sum_{m=n}^{\infty}\frac{e^{-N_{X}}}{n!(m-n)!}
    \left<
    1_{\mathcal{S}}, D
    \right>^{n}
    \left<
    1-1_{\mathcal{S}}, D
    \right>^{m-n}
    \label{eq:PoissonFovCardinality}
  \end{align}
  \label{pro:PoissonFovCardinality}
\end{proposition}
\textit{Proof:}
Substituting~(\ref{eq:PoissonDensity}) into~(\ref{eq:FovCardinalityGeneralExpanded}),
\begin{align}
  \rho_{\mathcal{S}}(n)
  &=
  \sum_{m=n}^{\infty}
  \frac{1}{m!}
  e^{-N_{X}}
  \int_{\mathbb{X}^{m}} \sum_{X^{n}\subseteq X}
  [1_{\mathcal{S}}(\cdot) D(\cdot)]^{X^{n}} \nonumber \\
  &\qquad \cdot
  [(1 - 1_{\mathcal{S}}(\cdot)) D(\cdot)]^{X \setminus X^{n}}
  \mathrm{d} \mathbf{x}_{1} \cdots \mathrm{d} \mathbf{x}_{m}
  \label{eq:PoissonFovCardinalityXmIntegral}
\end{align}
The nested integrals of~(\ref{eq:PoissonFovCardinalityXmIntegral}) can be distributed, rewriting the second sum over $n$-cardinality index sets $\mathcal{I}^{n}$ as
\begin{align}
  \rho_{\mathcal{S}}(n)
  &=
  \sum_{m=n}^{\infty}
  \frac{1}{m!}
  e^{-N_{X}}
  \, \,
  \sum_{\mathclap{\mathcal{I}^{n}\subseteq\mathbb{N}_{m}}} \quad
  \left[
    \int 1_{\mathcal{S}}(\mathbf{x}_{(\cdot)}) D(\mathbf{x}_{(\cdot)})\mathrm{d} \mathbf{x}_{(\cdot)}
  \right]^{\mathcal{I}^{n}} \nonumber \\
  &\qquad \cdot
  \left[
    \int(1 - 1_{\mathcal{S}}(\mathbf{x}_{(\cdot)})) D(\mathbf{x}_{(\cdot)})
  \right]^{\mathbb{N}_{m} \setminus \mathcal{I}^{n}}
\end{align}
Note that the value of the integrals is independent of the product index $i$, and thus
\begin{align}
  \rho_{\mathcal{S}}(n)
  &=
  \sum_{m=n}^{\infty}
  e^{-N_{X}}
  \frac{1}{m!}
  \frac{m!}{n!(m-n)!}
  \left<
    1_{\mathcal{S}}, D
  \right>^{n}
  \left<
    1 - 1_{\mathcal{S}}, D
  \right>^{m-n}
\end{align}
from which ~(\ref{eq:PoissonFovCardinality}) follows.
\null\hfill$\square$

\textit{Remark}: Computation of~(\ref{eq:PoissonFovCardinality}) requires only one integral computation; namely $\big<1_{\mathcal{S}}, D\big>$, which can be found either by summing the weights of~(\ref{eq:InSTimesPApproxGm}) or through Monte Carlo integration.
Using the integral property of the \ac{phd} (\ref{eq:PoissonIntegralIsExpectedCardinality}), the integral
\begin{align}
  \big<1 - 1_{\mathcal{S}}, D\big>=N_{X} - \big<1_{\mathcal{S}}, D\big>
\end{align}
Furthermore, for $m\gg N_{X}$, the summand of~(\ref{eq:PoissonFovCardinality}) is negligible, and the infinite sum can be safely truncated at an appropriately chosen $m=m_{\max}(N_{X})$.

\subsection{Independent Identically Distributed Cluster Distribution}
\label{ss:iidc}

The density of an \ac{iidc} \ac{rfs} is
\begin{align}
  f(X)
  =
  |X|!
  \cdot
  \rho(|X|)
  [p]^{X} \,,
  \label{eq:IidcDensity}
\end{align}
where $\rho(n)$ is the cardinality \ac{pmf} and $p(\mathbf{x})$ is the single-object state \ac{pdf}.

\begin{proposition}
  Given an \ac{iidc}-distributed \ac{rfs} with cardinality \ac{pmf} $\rho(\cdot)$ and state density $p(\cdot)$, the cardinality of objects inside the \ac{fov} $\mathcal{S}$ is distributed according to
  \begin{align}
    \rho_{\mathcal{S}}(n)
    =
    \sum_{m=n}^{\infty} \rho (m)
    \binom{m}{n}
    \big< 1_{\mathcal{S}}, p\big>^{n}
    \big< 1 - 1_{\mathcal{S}}, p\big>^{m-n}
    \label{eq:IidcFovCardinality}
  \end{align}
  where $\binom{m}{n}$ is the binomial coefficient.
  \label{pro:IidcFovCardinality}
\end{proposition}

\textit{Proof}:
Substituting~(\ref{eq:IidcDensity}) into~(\ref{eq:FovCardinalityGeneralExpanded}),
\begin{align}
  &\rho_{\mathcal{S}}(n) = \sum\limits_{m=n}^{\infty} \frac{1}{m!} m! \rho(m)\\
  &\quad \int\limits_{\mathbb{X}^{m}}
  \sum_{X^{n}\subseteq X}
  \cdot
  [1_{\mathcal{S}}(\cdot) p(\cdot)]^{X^{n}}
  [(1 - 1_{\mathcal{S}}(\cdot)) p(\cdot)]^{X \setminus X^{n}}
  \mathrm{d} \mathbf{x}_{1} \prodvarlist \mathrm{d} \mathbf{x}_{m}\nonumber
\end{align}
The integral can be moved inside the products so that
\begin{align}
  \rho_{\mathcal{S}}(n)
  =
  \sum\limits_{m=n}^{\infty} \rho(m)
  \sum_{\mathcal{I}^{n}\subseteq \mathbb{N}_{m}}
  \left[
    \int 1_{\mathcal{S}}(\mathbf{x}_{(\cdot)}) p(\mathbf{x}_{(\cdot)}) \mathrm{d} \mathbf{x}_{(\cdot)}
  \right]^{\mathcal{I}^{n}} \nonumber\\
  \cdot
  \left[
    \int (1 - 1_{\mathcal{S}}(\mathbf{x}_{(\cdot)})) p(\mathbf{x}_{(\cdot)}) \mathrm{d} \mathbf{x}_{(\cdot)}
  \right]^{\mathbb{N}_{m} \setminus \mathcal{I}^{n}}
  \label{eq:IidcFovCardinalityIndexSet}
\end{align}
Equation~(\ref{eq:IidcFovCardinality}) follows from~(\ref{eq:IidcFovCardinalityIndexSet}) by noting that there are $\binom{m}{n}$ unique unordered $n$-cardinality index subsets of $\mathbb{N}_{m}$.
\null\hfill$\square$

\subsection{Multi-Bernoulli Distribution}
\label{ss:mb}
The density of a \ac{mb} distribution is \cite[p.\ 102]{MahlerAdvancesStatisticalMultitargetFusion14}
\begin{align}
  f(X)
  =
  \left[
    \left(
      1 - r^{(\cdot)}
    \right)
  \right]^{\mathbb{N}_{M}}
  \sum\limits_{\mathclap{\vphantom{\big[}1 \leq i_{1} \neq \cdots \neq i_{n} \leq M}}
    \quad
  \left[
    \dfrac{
      r^{i_{(\cdot)}} p^{i_{(\cdot)}}(\mathbf{x}_{(\cdot)})
    }{
      1 - r^{i_{(\cdot)}}
    }
  \right]^{\mathrlap{\mathbb{N}_{n}}}
  \label{eq:MbDensity}
\end{align}
where $M$ is the number of \ac{mb} components and maximum possible object cardinality, $r^{i}$ is the probability that the $i$\textsuperscript{th} object exists, and $p^{i}(\mathbf{x})$ is the single-object state density of the $i$\textsuperscript{th} object if it exists.

\begin{proposition}
\label{pro:MbFovCardinality}%
  Given at \ac{mb} density of the form of~(\ref{eq:MbDensity}), the cardinality of objects inside the \ac{fov} $\mathcal{S}$ is distributed according to
  \begin{align}
    &\rho_{\mathcal{S}}(n) =
    \left[
      \left(
        1 - r^{(\cdot)}
      \right)
    \right]^{\mathbb{N}_{M}} \nonumber \\
    &\quad
    \cdot
    \sum\limits_{\mathclap{\mathcal{I}_{1} \uplus \mathcal{I}_{2} \subseteq \mathbb{N}_{M}}}
    \delta_{n}(|\mathcal{I}_{1}|)
    \left[
      \dfrac{%
        \left< 1_{\mathcal{S}}, r^{(\cdot)} p^{(\cdot)}\right>
      }{%
        1 - r^{(\cdot)}
      }
    \right]^{\mathcal{I}_{1}}
    \left[
      \dfrac{%
        \left<1 - 1_{\mathcal{S}},
        r^{(\cdot)} p^{(\cdot)}\right>
      }{%
        1 - r^{(\cdot)}
      }
    \right]^{\mathcal{I}_{2}}
    \label{eq:MbFovCardinality}
  \end{align}
  where the summation is taken over all mutually exclusive index partitions $\mathcal{I}_{1}$, $\mathcal{I}_{2}$ such that $\mathcal{I}_{1} \cup \mathcal{I}_{2} \subseteq \mathbb{N}_{M}$.
\end{proposition}

Proof of Proposition~\ref{pro:MbFovCardinality} is given in Appendix~\ref{a:ProofOfMbFovCardinality}.
Within a given summand term of (\ref{eq:MbFovCardinality}), the index sets $\mathcal{I}_{1}$, $\mathcal{I}_{2}$, and $\mathbb{N}_{M}\setminus(\mathcal{I}_{1}\cup\mathcal{I}_{2})$ can be interpreted as the indices of objects within the \ac{fov}, objects outside the \ac{fov}, and non-existent objects, respectively.
Following the same procedure, similar results for the \ac{lmb} \cite{ReuterLmbFilter14} and \ac{mbm} \cite{WilliamsMarginalMultiBernoulliMixturePGFL15} \ac{rfs} distributions may be obtained.

Direct computation of~(\ref{eq:MbFovCardinality}) is only feasible for small $M$ due to the sum over all permutations $\mathcal{I}_{1} \uplus \mathcal{I}_{2} \subseteq \mathbb{N}_{M}$.
For large $M$, an alternative formulation based on Fourier transforms allows fast numerical computation.
For each \ac{mb} component, the integral $\left<1_{\mathcal{S}}, p^{(i)}\right>$ is computed either by summing the weights of the partitioned \ac{gm} or by Monte Carlo integration.
Using the integral results, the probability of object $i$ existing inside the \ac{fov} is found as
\begin{align}
  r_{\mathcal{S}}^{(i)} = r^{(i)}\left< 1_{\mathcal{S}},p^{(i)}\right>
\end{align}
Then, as shown in \cite{FernandezPoissonBinomial10}, (\ref{eq:MbFovCardinality}) can be equivalently written as
\begin{align}
  \label{eq:MbFovCardinalityFourier}
  &\rho_{\mathcal{S}}(n)
  =
  \frac{1}{M+1} \times\\
  &\sum_{m=0}^{M}
  \left\lbrace
    e^{-j2\pi m n /(M + 1)}
    \prod_{k=1}^{M}
    \left[
      r_{\mathcal{S}}^{(k)}
      e^{j2\pi m/(M+1)}
      +
      (1-r_{\mathcal{S}}^{(k)})
    \right]
  \right\rbrace
  \nonumber
\end{align}
and solved using the discrete Fourier transform, for which a number of efficient algorithms exist.
\subsection{Generalized Labeled Multi-Bernoulli Distribution}
\label{ss:glmb}
The density of a \ac{glmb} distribution is given by \cite{VoLabeledRfsGlmbFilter14}
\begin{align}
  \label{eq:GlmbDensity}
  \mathring{f} (\mathring{X})
  =
  \Delta (\mathring{X})
  \sum\limits_{\xi \in \Xi}
  w^{(\xi)} (\mathcal{L} (\mathring{X}))
  [p^{(\xi)}]^{\mathring{X}} \,,
\end{align}
where each $\xi\in\Xi$ represents a history of measurement association maps, each $p^{(\xi)}(\cdot,\ell)$ is a probability density on $\mathbb{X}$, and each weight $w^{(\xi)}$ is non-negative with ${\sum\limits_{(I,\xi)\in \mathcal{F}(\mathbb{L})\times\Xi} w^{(\xi)}(I)=1}$.  The label of a labeled state $\mathring{x}$ is recovered by $\mathcal{L}(\mathring{x})$, where $\mathcal{L} : \mathbb{X} \times \mathbb{L} \mapsto \mathbb{L}$ is the projection defined by $\mathcal{L}((\mathbf{x}, \ell)) \triangleq \ell$.  Similarly, for \acp{lrfs}, $\mathcal{L}(\mathring{X}) \triangleq \{\mathcal{L}(\mathring{x}) : \mathring{x} \in \mathring{X}\}$.  The distinct label indicator  $\Delta(\mathring{X})=\delta_{(|\mathring{X}|)}(|\mathcal{L}(\mathring{X})|)$ ensures that only sets with distinct labels are considered.

\begin{proposition}
  Given a \ac{glmb} density $\mathring{f} (\mathring{X})$ of the form of~(\ref{eq:GlmbDensity}), the cardinality of objects inside a bounded \ac{fov} $\mathcal{S}$ is distributed according to
\begin{align}
  \label{eq:GlmbFovCardinality}
  \rho_{\mathcal{S}}(n)
  =&
  \sum\limits_{\mathclap{(\xi, \mathcal{I}_{1} \uplus \mathcal{I}_{2}) \in \Xi \times \mathcal{F}(\mathbb{L})}}
  w^{(\xi)}(I)
  \delta_{n}(|\mathcal{I}_{1}|)
  \left<1_{\mathcal{S}}, p\right>^{\mathcal{I}_{1}}
  \left<1-1_{\mathcal{S}}, p\right> ^{\mathcal{I}_{2}}
\end{align}
\end{proposition}

\textit{Proof}:
Equation~(\ref{eq:ConditionalCardinality}) can be rewritten to accommodate the labeled \ac{rfs} as
\begin{align}
  \rho_{\mathcal{S}}(n \,|\, \mathring{X}) = \sum_{\mathring{X}^{n} \subseteq \mathring{X}} [1_{\mathcal{S}}(\cdot)]^{\mathring{X}^{n}} [1 - 1_{\mathcal{S}}(\cdot)]^{\mathring{X}\setminus \mathring{X}^{n}}
  \label{eq:ConditionalCardinalityLabeled}
\end{align}
If $\mathring{X}$ is distributed according to the \ac{lrfs} density $\mathring{f}(\mathring{X})$, the \ac{fov} cardinality distribution is obtained via the set integral
\begin{align}
  \rho_{\mathcal{S}}(n) = \int \rho_{\mathcal{S}}(n\,|\,\mathring{X}) \mathring{f}(\mathring{X}) \delta \mathring{X}
\end{align}
Expanding the integral,
\begin{align}
  &\rho_{\mathcal{S}}(n) \nonumber \\
  &=
  \sum_{m=n}^{\infty} \frac{1}{m!}
  \sum_{(\ell_{1} \varlist \ell_{m}) \in
  \mathbb{L}^{m}} \,
  \int\limits_{\mathclap{\mathbb{X}^{m}}}
  \rho_{\mathcal{S}}(n\,|\,\{(\mathbf{x}_{1},\ell_{1}) \varlist (\mathbf{x}_{m},\ell_{m})\}) \nonumber \\
  &\qquad\cdot
  \mathring{f}(\{(\mathbf{x}_{1},\ell_{1}) \varlist (\mathbf{x}_{m},\ell_{m})\}) \mathrm{d} \mathbf{x}_{1} \cdots \mathrm{d} \mathbf{x}_{m}
\end{align}
Defining $p^{(\xi,\ell)}(x)\triangleq p^{(\xi)}(x,\ell)$, substitution of~(\ref{eq:GlmbDensity}) and~(\ref{eq:ConditionalCardinalityLabeled}) yields
\begin{align}
  \rho_{\mathcal{S}}(n)
  &=
  \sum_{m=n}^{\infty} \frac{1}{m!}
  m!
  \sum_{\{\ell_{1},\dots,\ell_{m}\} \in \mathbb{L}^{m}}
  \sum_{\xi \in \Xi}
  w^{(\xi)}(\{\ell_{1}, \dots, \ell_{m}\}) \nonumber \\
  &\qquad
  \sum_{\mathclap{I^{n} \subseteq \{\ell_{1}, \dots \ell_{m}\}}}
  \quad
  \big<1_{\mathcal{S}},p^{(\xi,\cdot)}\big>^{I^{n}}
  \big<1 - 1_{\mathcal{S}},p^{(\xi,\cdot)}\big>^{\{\ell_{1}, \dots, \ell_{m}\} \setminus I^{n}} \nonumber \\[1em]
  &=
  \quad
  \sum_{\mathclap{(\xi, I) \in \Xi \times \mathcal{F}(\mathbb{L})}}
  w^{(\xi)}(I)
  \sum_{I^{n} \subseteq I}
  \big<1_{\mathcal{S}},p^{(\xi,\cdot)}\big>^{I^{n}}
  \big<1 - 1_{\mathcal{S}},p^{(\xi,\cdot)}\big>^{I \setminus I^{n}}
\end{align}
from which~(\ref{eq:GlmbFovCardinality}) follows.
\null\hfill$\square$

\textit{Remark}: Substitution of $n=0$ in \ref{eq:GlmbFovCardinality} gives the \ac{glmb} void probability functional \cite[Eq. 22]{BeardVoidProbabilitiesCauchySchwarzGlmb17}, which, while less general, has theoretical significance and practical applications in sensor management.
\section{Sensor Placement Example}
\label{sec:SensorPlacementExample}
\begin{figure}[htbp]
  \centering
  \includegraphics[width=\linewidth]{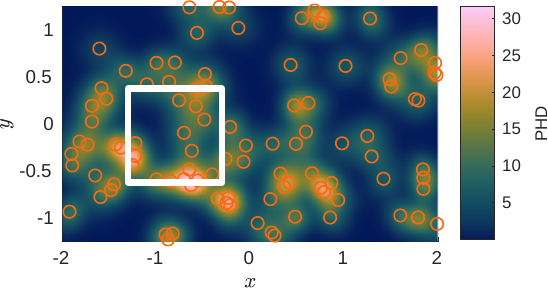}
  \caption{PHD of the global MB distribution with 100 potential objects, where object means are represented by orange circles and the bounds of the FoV that maximizes the FoV cardinality variance are shown in white.}
  \label{fig:ExampleMaxCardVarPhd}
\end{figure}

The \ac{fov} statistics developed in this paper are demonstrated through a sensor placement optimization problem subject to multi-object uncertainty.  The global distribution is assumed to be \ac{mb}-distributed. Numerical simulation is performed for the case of $100$ \ac{mb} components, with probabilities of existence randomly chosen between $0.35$ and $1$.  Each \ac{mb} component has a Gaussian density and randomly chosen mean and covariance.  To visualize the global distribution, the \ac{phd} is shown in Figure~\ref{fig:ExampleMaxCardVarPhd}.

The \ac{phd} is analogous to the expected value for \acp{rfs} and is defined as \cite{mahler2007phd}
\begin{align}
  D(\mathbf{x}) \triangleq \mathrm{E}[\delta_{X}(\mathbf{x})] = \int \delta_{X}(\mathbf{x}) \cdot f(X)\delta X
\end{align}
for an arbitrary \ac{rfs} $X$ with density $f(X)$, where
\begin{align}
  \delta_{X}(\mathbf{x}) \triangleq \sum_{\mathbf{w}\in X} \delta_{\mathbf{w}}(\mathbf{x})
\end{align}
It follows that the \ac{phd} of an \ac{mb} \ac{rfs} (\ref{eq:MbDensity}) is \cite[p.\ 102]{MahlerAdvancesStatisticalMultitargetFusion14}
\begin{align}
  D(\mathbf{x}) = \sum_{i=1}^{M} r^{i}p^{i}(\mathbf{x})
\end{align}

\begin{figure}[htbp]
  \centering
  \includegraphics[width=\linewidth]{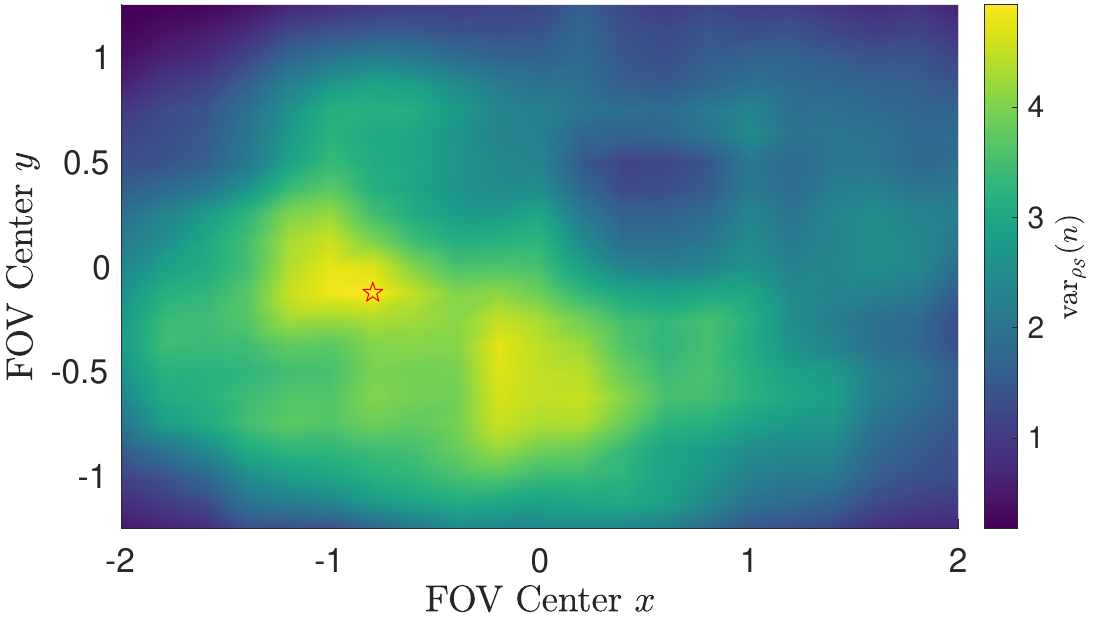}
  \caption{FoV cardinality variance as a function of FoV center location, where the red star denotes the maximum variance point.}
  \label{fig:ExampleMaxCardVarVar}
\end{figure}

The objective of the sensor control problem is to place the \ac{fov}, comprised of a square of $1\times 1$ dimensions, in the \ac{roi} (Figure \ref{fig:ExampleMaxCardVarPhd}) such that the variance of object cardinality inside the \ac{fov} is maximized.  This objective can be interpreted as placing the \ac{fov} in a region of the \ac{roi} where the object cardinality is most uncertain.  A related objective which minimizes the variance of the \textit{global} cardinality using CB-MeMBer predictions was first proposed in \cite{HoangSensorManagementMultiBernoulliRenyiMaxCardinalityVariance14}. For each candidate \ac{fov} placement, the \ac{fov} cardinality \ac{pmf} is given by~(\ref{eq:MbFovCardinality}) and is efficiently computed using (\ref{eq:MbFovCardinalityFourier}).  The variance of the resulting \ac{pmf} is shown as a function of the \ac{fov} center location in Figure~\ref{fig:ExampleMaxCardVarVar}.  The optimal \ac{fov} center location is found to be $(-0.8, -1.25)$.

A compelling result is that, by virtue of the bounded \ac{fov} geometry, spatial information is encoded in the \ac{fov} cardinality \ac{pmf}.  It can be seen that the optimal \ac{fov} (Fig.~\ref{fig:ExampleMaxCardVarPhd}) has boundary segments (lower half of left boundary and right half of lower boundary) that bisect clusters of \ac{mb} components.  These boundary segments divide the components' single-object densities such that significant mass appears inside and outside the \ac{fov}, increasing the overall \ac{fov} cardinality variance.

\section{Conclusions}
\label{sec:Conclusions}
This paper presents an approach for incorporating bounded \acf{fov} geometry into state density updates and object cardinality predictions via finite set statistics.
Inclusion/exclusion evidence such as negative information and soft evidence is processed in state density updates via a novel Gaussian splitting algorithm that recursively refines a \acl{gm} approximation near the boundaries of the discrete \ac{fov} geometry.  Using \acl{fisst}, cardinality probability mass functions that describe the probability that a given number of objects exist inside the \ac{fov} are derived.  The approach is presented for representative labeled and unlabeled random finite set  distributions and, thus, is applicable to a wide range of tracking, perception, and sensor planning problems.
\bibliographystyle{IEEEtran}
\bibliography{keithlegrand_refs,zotero_refs}

\appendices
\section{Inclusion Consistency Example}%
\label{sec:inclusion_consistency_example}

\begin{figure}[htpb]
  \centering
  \ifx\undefined\usetikz
  \includegraphics{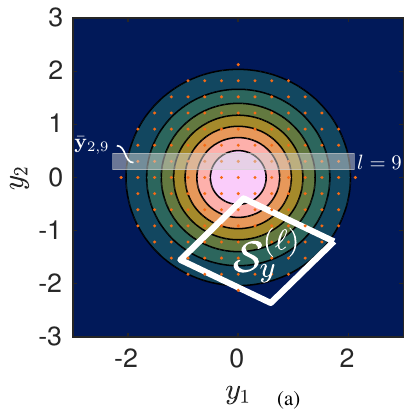}
  \else
      \input{figures/inclusion_variable_explanation.tex}
  \fi
     \caption{(a) True trajectory and state estimates over time, where position state densities are shown for time steps $k=15,25,55$ ($t=225,375,825\,[\textrm{s}]$) and (b)~posterior probability of existence over time.}%
      \label{fig:InclusionVariableExplanation}
\end{figure}

  Consider a plane of constant $y_{2}=\bar{y}_{2}(9)$--that is, $j=2$ and $l=9$.
  As pictured in Fig.~\ref{fig:InclusionVariableExplanation}, the index $l=9$ denotes the ninth grid plane from the bottom.
  To evaluate inclusion/exclusion consistency in this plane, an arbitrary reference point is selected as $\bar{\mathbf{y}}_{2,9}$ (where the corresponding indices are $i_{1}'=2$ and $i_{j}=i_{2}=l=9$).
  Note that this reference index tuple $(2,9)$ belongs to $G$ (depicted by the set of orange dots) and lies in the plane of constant $i_{j}=l$.

  It is apparent from Fig.~\ref{fig:InclusionVariableExplanation} that $\bar{\mathbf{y}}_{2,9}\notin\mathcal{S}_{y}^{(\ell)}$.
  Thus, the corresponding component inclusion variable (\ref{eq:ComponentInclusionVariable}) for the selected reference point is
\begin{align}
  \label{eq:ComponentInclusionVariableExample}
  d_{i_{1}',i_{2}}^{(\ell)}
  =
  d_{2,9}^{(\ell)}
  =
  1_{\mathcal{S}_{y}^{(\ell)}}(\bar{\mathbf{y}}_{2, 9}) =0
\end{align}
In the following inclusion/exclusion consistency check, which follows from
(\ref{eq:ConsistentPlaneIndicator}), the inclusion variables are computed for all remaining points in the plane and compared to $d_{2,9}$:
\begin{align}
  s_{j}^{(\ell)}(l)=s_{2}^{(\ell)}(9)
  &=
  \prod_{G, i_{2}=9}
  \delta_{d_{2,9}^{(\ell)}}(d_{i_{1},9}^{(\ell)})\nonumber\\
  &=
  \delta_{d_{2,9}^{(\ell)}}(d_{2,9}^{(\ell)})
  \cdot
  \delta_{d_{2,9}^{(\ell)}}(d_{3,9}^{(\ell)})
  \cdots
  \delta_{d_{2,9}^{(\ell)}}(d_{14,9}^{(\ell)})\nonumber\\
  &=
  \delta_{0}(0)\cdot \delta_{0}(0) \cdots \delta_{0}(0)= 1
\end{align}
where it is noted that $i_{1}$ ranges from two to fourteen in the considered plane (in which there are thirteen corresponding orange dots).
Thus, $s_j^{(\ell)}(l)=1$ signifies that the plane is indeed consistently inside or consistently outside the \ac{fov}, the latter of which is easily verified by inspecting Fig.~\ref{fig:InclusionVariableExplanation}.

\section{Taylor Series Expansion About Means}%
\label{a:change1_taylor_series_expansion_about_means}

Equation~(\ref{eq:BernoulliStatePdfUpdate}) can be written compactly as
\begin{align}
  p_{k|k}(\mathbf{x}) &= \alpha(\mathbf{x}) p_{k|k-1}(\mathbf{x})\\
  &=\sum_{\ell=1}^{L_{k|k-1}}
  \alpha(\mathbf{x})w_{k|k-1}^{(\ell)}
  \mathcal{N}(\mathbf{x}; \, \mathbf{m}_{k|k-1}^{(\ell)}, \, \mathbf{P}_{k|k-1}^{(\ell)})
\end{align}
where
\begin{align}
  \alpha(\mathbf{x}) =
  \frac{1-p_{D}(\mathbf{x}; \mathcal{S}_{k}) + p_{D}(\mathbf{x}; \mathcal{S}_{k}) \sum\limits_{\boldsymbol{\zeta} \in \Upsilon_{k}} \frac{\tilde{g}_{k}(\boldsymbol{\zeta} | \mathbf{x})}{\lambda_{c} \tilde{c}(\boldsymbol{\zeta})}}{1 - \Delta_k}
\end{align}
and where the functional dependence of $\alpha$ on the \ac{fov} and measurement is omitted for brevity.
The function $\alpha(\mathbf{x})$ can be approximated locally by a Taylor series expansion about a given component mean as
\begin{align}
  \alpha(\mathbf{x}) \approx \alpha(\mathbf{m}_{k|k-1}^{(\ell)}) +\left. \left(\frac{\partial \alpha}{\partial \mathbf{x}}\right)\right|_{\mathbf{x}=\mathbf{m}_{k|k-1}^{(\ell)}}(\mathbf{x}-\mathbf{m}_{k|k-1}^{(\ell)}) + \cdots
\end{align}
To zeroth order, $\alpha(\mathbf{x}) \approx \alpha(\mathbf{m}_{k|k-1}^{(\ell)})$, such that
\begin{align}
  p_{k|k}(\mathbf{x}) \approx
  \sum_{\ell=1}^{L_{k|k-1}}
  \alpha(\mathbf{m}_{k|k-1}^{(\ell)})w_{k|k-1}^{(\ell)}
  \mathcal{N}(\mathbf{x}; \, \mathbf{m}_{k|k-1}^{(\ell)}, \, \mathbf{P}_{k|k-1}^{(\ell)})
\end{align}
from which (\ref{eq:GmBernoulliUpdatedMixture})-(\ref{eq:GmBernoulliUpdatedCov}) follow.

\section{Proof of Proposition~\ref{pro:MbFovCardinality}}
\label{a:ProofOfMbFovCardinality}
Let $\mathbb{K}_{M}^{(n)} \triangleq \{(i_{1}\varlist i_{n}) \, : \, 1\leq i_{1} \neq \cdots \neq i_{n} \leq M\}$.
Then,~(\ref{eq:MbDensity}) can be rewritten as
\begin{align}
  f(X)
  =
  \left[
    \left(
      1 - r^{(\cdot)}
    \right)
  \right]^{\mathbb{N}_{M}}
  \sum\limits_{(\mathcal{I}_{\sigma}) \in \mathbb{K}_{M}^{(n)}}
  \left[
    \dfrac{%
      r^{i_{(\cdot)}} p^{i_{(\cdot)}}(x_{(\cdot)})
    }{%
      1 - r^{i_{(\cdot)}}
    }
  \right]^{\mathbb{N}_{n}}
  \label{eq:MbDensityIndexSet}
\end{align}
 where $\mathcal{I}_{\sigma}$ denotes the (unordered) set $\{i_{1} \varlist i_{n}\}$ and~$(\mathcal{I}_{\sigma})$ denotes the (ordered) sequence $(i_{1} \varlist i_{n}) = (\alpha_{\sigma(1)}\varlist \alpha_{\sigma(n)})$, where the $n$-tuple index set $\{\alpha_{1} \varlist \alpha_{n}\} \subseteq \mathbb{N}_{M}$ and $\sigma$ is a permutation of $\mathbb{N}_{n}$.

Substituting~(\ref{eq:MbDensityIndexSet}) into~(\ref{eq:FovCardinalityGeneralExpanded}),
\begin{align}
  &\rho_{\mathcal{S}}(n)
  =
  \left[
    \left(
      1 - r^{(\cdot)}
    \right)
  \right]^{\mathbb{N}_{M}} \nonumber \\
  &\quad
  \cdot
  \sum_{m=n}^{M}
  \frac{1}{m!}
  \int_{\mathbb{X}^{m}}
  \sum\limits_{(\mathcal{I}_{\sigma}) \in \mathbb{K}_{M}^{(n)}}
  \delta_{m}(|\mathcal{I}_{\sigma}|)
  \left[
    \dfrac{%
      r^{i_{(\cdot)}} p^{i_{(\cdot)}}(\mathbf{x}_{(\cdot)})
    }{%
      1 - r^{i_{(\cdot)}}
    }
  \right]^{\mathbb{N}_{m}}\nonumber\\
  &\qquad
  \sum_{X^{n}\subseteq X} [1_{\mathcal{S}}(\cdot)]^{X^{n}} [1 - 1_{\mathcal{S}}(\cdot)]^{X \setminus X^{n}}
  \mathrm{d} \mathbf{x}_{1} \cdots \mathrm{d} \mathbf{x}_{m}
\end{align}
The last sum can be written in terms of label index sets $\mathcal{I}_{1}\uplus \mathcal{I}_{2}=\mathcal{I}_{\sigma}$ as
\begin{align}
  &\rho_{\mathcal{S}}(n)
  =
  \left[
    \left(
      1 - r^{(\cdot)}
    \right)
  \right]^{\mathbb{N}_{M}} \\
  &\quad
  \cdot
  \sum_{m=n}^{M}
  \frac{1}{m!}
  \int_{\mathbb{X}^{m}}
  \sum\limits_{(\mathcal{I}_{\sigma}) \in \mathbb{K}_{M}^{(n)}}
  \delta_{m}(|\mathcal{I}_{\sigma}|)
  \left[
    \dfrac{%
      r^{i_{(\cdot)}} p^{i_{(\cdot)}}(\mathbf{x}_{(\cdot)})
    }{%
      1 - r^{i_{(\cdot)}}
    }
  \right]^{\mathbb{N}_{m}}\nonumber\\
  &\quad
  \cdot
  \sum_{\mathclap{\mathcal{I}_{1} \uplus \mathcal{I}_{2}=\mathcal{I}_{\sigma}}}
  \delta_{n}(|\mathcal{I}_{1}|)
  [1_{\mathcal{S}}(\mathbf{x}_{(\cdot)})]^{\{j:i_{j}\in\mathcal{I}_{1}\}}
  [1 - 1_{\mathcal{S}}(\mathbf{x}_{(\cdot)})]^{\{j:i_{j}\in \mathcal{I}_{2}\}} \nonumber\\
  &\qquad \,
  \mathrm{d} \mathbf{x}_{1} \cdots \mathrm{d} \mathbf{x}_{m} \nonumber
\end{align}
where the innermost sum is taken over all mutually disjoint subsets $\mathcal{I}_{1}$, $\mathcal{I}_{2}$ such that $\mathcal{I}_{1}\cup \mathcal{I}_{2}=\mathcal{I}_{\sigma}$.
Distributing terms from the second summation,
\begin{align}
  &\rho_{\mathcal{S}}(n)
  =
  \left[
    \left(
      1 - r^{(\cdot)}
    \right)
  \right]^{\mathbb{N}_{M}} \\
  &\quad
  \cdot
  \sum_{m=n}^{M}
  \frac{1}{m!}
  \int_{\mathbb{X}^{m}}
  \sum\limits_{(\mathcal{I}_{\sigma}) \in \mathbb{K}_{M}^{(n)}}
  \delta_{m}(|\mathcal{I}_{\sigma}|)
  \sum_{\mathcal{I}_{1} \uplus \mathcal{I}_{2}=\mathcal{I}_{\sigma}} \delta_{n}(|\mathcal{I}_{1}|) \nonumber\\
  &\quad
  \cdot
  \left[
    \dfrac{%
    1_{\mathcal{S}}(\mathbf{x}_{(\cdot)})r^{i_{(\cdot)}} p^{i_{(\cdot)}}(\mathbf{x}_{(\cdot)})
    }{%
      1 - r^{i_{(\cdot)}}
    }
  \right]^{\{j:i_{j}\in\mathcal{I}_{1}\}} \nonumber \\
  &\quad
  \cdot
  \left[
    \dfrac{%
      [1 - 1_{\mathcal{S}}(\mathbf{x}_{(\cdot)})]
      r^{i_{(\cdot)}} p^{i_{(\cdot)}}(\mathbf{x}_{(\cdot)})
    }{%
      1 - r^{i_{(\cdot)}}
    }
  \right]^{\{j:i_{j}\in \mathcal{I}_{2}\}}
  \mathrm{d} \mathbf{x}_{1} \cdots \mathrm{d} \mathbf{x}_{m} \nonumber
\end{align}
Because $\mathcal{I}_{1} \cap \mathcal{I}_{2}= \emptyset$, then $\{\mathbf{x}_{j} : i_{j}\in \mathcal{I}_{1} \} \cap \{ \mathbf{x}_{j} : i_{j}\in \mathcal{I}_{2} \} = \emptyset $ and the integral on $\mathbb{X}^{m}$ becomes a product of integrals on $\mathbb{X}$, such that
\begin{align}
  &\rho_{\mathcal{S}}(n)
  =
  \left[
    \left(
      1 - r^{(\cdot)}
    \right)
  \right]^{\mathbb{N}_{M}} \\
  &\quad
  \cdot
  \sum_{m=n}^{M}
  \frac{1}{m!}
  \sum\limits_{(\mathcal{I}_{\sigma}) \in \mathbb{K}_{M}^{(n)}}
  \delta_{m}(|\mathcal{I}_{\sigma}|)
  \sum_{\mathcal{I}_{1} \uplus \mathcal{I}_{2}=\mathcal{I}_{\sigma}} \delta_{n}(|\mathcal{I}_{1}|) \nonumber\\
  \qquad &
  \cdot
  \left[
    \dfrac{%
      \left< 1_{\mathcal{S}}, r^{i_{(\cdot)}} p^{i_{(\cdot)}}\right>
    }{%
      1 - r^{i_{(\cdot)}}
    }
  \right]^{\{j:i_{j}\in\mathcal{I}_{1}\}}
  \left[
    \dfrac{%
      \left<1 - 1_{\mathcal{S}},
      r^{i_{(\cdot)}} p^{i_{(\cdot)}}\right>
    }{%
      1 - r^{i_{(\cdot)}}
    }
  \right]^{\{j:i_{j}\in \mathcal{I}_{2}\}} \nonumber
\end{align}
Now note that the result of the innermost sum does not depend the permutation order of $(\mathcal{I}_{\sigma})$.  Thus the property \cite[Lemma 12]{VoLabeledRfsConjugatePriors13}, which states that for an arbitrary symmetric function $h$,
\begin{align}
  \sum_{(i_{1}, \ldots, i_{m})} h(\{i_{1}, \ldots, i_{m}\})
  = m!
  \sum_{\{i_{1}, \ldots, i_{m}\}} h(\{i_{1}, \ldots, i_{m}\})
\end{align}
is applied, yielding
\begin{align}
  &\rho_{\mathcal{S}}(n)
  =
  \left[
    \left(
      1 - r^{(\cdot)}
    \right)
  \right]^{\mathbb{N}_{M}}  \\
  &\quad
  \cdot
  \sum_{m=n}^{M}
  \sum\limits_{\mathcal{I}_{1} \uplus \mathcal{I}_{2} \subseteq \mathbb{N}_{M}}
  \delta_{m}(|\mathcal{I}_{1}\uplus \mathcal{I}_{2}|)
  \delta_{n}(|\mathcal{I}_{1}|) \nonumber \\
  &\quad
  \cdot
  \left[
    \dfrac{%
      \left< 1_{\mathcal{S}}, r^{(\cdot)} p^{(\cdot)}\right>
    }{%
      1 - r^{(\cdot)}
    }
  \right]^{\mathcal{I}_{1}}
  \left[
    \dfrac{%
      \left<1 - 1_{\mathcal{S}},
      r^{(\cdot)} p^{(\cdot)}\right>
    }{%
      1 - r^{(\cdot)}
    }
  \right]^{\mathcal{I}_{2}} \nonumber
\end{align}
The term $\delta_{m}(|\mathcal{I}_{1} \uplus \mathcal{I}_{2}|)$ is non-zero only when the combined cardinality of $\mathcal{I}_{1}$ and $\mathcal{I}_{2}$ is equal to $m$---the index of the outermost sum.  Thus, the outermost sum is absorbed by the second sum to give~(\ref{eq:MbFovCardinality}).
\null\hfill$\square$

\end{document}